\documentclass[11pt]{article}
\usepackage{color}
\usepackage{amsmath}
\usepackage{amssymb}
\usepackage{latexsym}
\usepackage{epsf}
\usepackage[greek,french,english]{babel}
\usepackage[utf8x]{inputenc}
\usepackage{amsfonts}
\usepackage{yfonts}
\usepackage{bm}
\usepackage{graphicx}
\usepackage{physics}
\def\im{\mathrm {i}}

\def\IR{\relax{\rm I\kern-.18em R}}
\def\I1{\relax{\rm 1\kern-.40em 1}}
\def\IZ{\relax{\rm Z\kern-.40em Z}}
\def\be{\begin{equation}}
\def\ee{\end{equation}} 
\begin{document}
\begin{center}
{\bf \Large {LA PHYSIQUE TH\'EORIQUE DES HAUTES 
 \vskip 0.3cm
 \'ENERGIES EN FRANCE}}
\vskip 0.3 cm
{\bf De la Mécanique Quantique au Modèle Standard}
\vskip 0.3 cm
{\bf A l'occasion du 150e anniversaire de la Société Française de Physique}
\vskip 0.5cm
Jean Iliopoulos -- \'Ecole Normale Supérieure
\vskip 0.3cm
Paris, juillet 2023
\vskip 1cm
{\bf ABSTRACT}
\end{center}
The year 2023 marks three anniversaries: the 150th anniversary of the French Physical Society, the 100th anniversary since the  publication of Louis de Broglie's articles -- the first fundamental articles of Quantum Mechanics -- and the 50th anniversary  of the Standard Model, the theory which describes in a unified framework the strong, the electromagnetic and the weak interactions among elementary particles. This note is an attempt to present the situation of theoretical physics in France during these years and highlight the french contributions in this world effort. We shall see that, although these contributions have been important, french theorists rarely played the leading roles. At the end I will try to draw some conclusions from these findings.
\vskip 0.5cm
\section{Avant-propos} 

L'année 2023 marque trois anniversaires: le 150e anniversaire de la création de la Société Française de Physique (SFP), le 100e anniversaire de la publication des articles de Louis de Broglie\footnote{ Louis Victor, Prince, puis Duc, de Broglie (1892-1987)}, premiers travaux fondateurs de la nouvelle mécanique quantique et le 50e anniversaire de la construction du Modèle Standard, la théorie qui décrit les interactions faibles, électromagnétiques et fortes entre les particules élémentaires. Durant ces 150 ans la physique a fait des progrès spectaculaires de sorte que, aujourd'hui, elle est omniprésente dans notre vie. La SFP a voulu présenter ces progrès  dans un livre destiné à un public plus large\cite{LivreSFP}. Cette note est une version un peu différente de ma contribution à ce livre. Je veux montrer que les théoriciens français de la physique des hautes énergies ont contribué de façon significative à ce progrès, même si ils ont rarement joué les premiers rôles. 
Il ne s'agit pas d'un article sur l'histoire de la physique parce que je me limiterai à la physique théorique des particules élémentaires et, même dans ce cadre, je vais mentionner uniquement les étapes les plus importantes dans la voie vers le Modèle Standard et citer explicitement les seules contributions françaises. Néanmoins, c'est un article d'histoire écrit par quelqu'un qui n'est pas un historien. J'ai essayé d'être prudent et vérifier, dans la mesure du possible, mes sources, mais je ne suis pas sûr d'avoir échappé à tous les pièges\footnote{R.P. Feynman a remarqué que les scientifiques se ridiculisent (“make fools of themselves”) chaque fois qu'ils s'expriment en dehors de leur spécialité. L'histoire des sciences n'étant pas ma spécialité, cet article risque de montrer que, encore une fois, Feynman avait raison.}. D'un autre côté il s'agit d'une histoire passionnante qui mérite d'être étudiée en détail, ne serait-ce que parce qu'elle peut nous montrer des failles possibles dans notre système d'éducation. J'espère que cette note pourra servir de motivation à un véritable historien des sciences pour se lancer dans cette étude.

Pendant la rédaction de cette note j'ai bénéficié de conseils de plusieurs collègues. En particulier, je tiens à remercier MM Yves Sacquin et Daniel Treille pour leurs nombreuses remarques et corrections.

\section{Introduction}

La  division “Champs et Particules” de la SFP fut créée en 1969 par séparation à partir de la Division de Physique Nucléaire. Son titre est imprécis et on devrait dire plutôt “Champs {\it Quantiques} et Particules {\it Elémentaires}”. 

Si la nature discontinue de la matière a fait l'objet des grands débats scientifiques jusqu'à la fin du 19e siècle\footnote{On attribue à Démocrite d'Abdère, philosophe Grec (460 av. J.-C. - 370 av. J.-C.) la paternité de l'hypothèse d'une structure discontinue de la matière. Pour Démocrite la matière est composée d'“atomes” et du “vide”. \textgreek{“Νόμω γάρ χροιή, νόμω γλυκύ, νόμω πικρόν, ετεή δ' άτομα καί κενόν"}, ou “des lois déterminent la couleur, le doux ou l'amer, mais tout est composé d'atomes et du vide”.}, aujourd'hui la question est close. A l'aide de “microscopes” de plus en plus puissants (en particulier des accélérateurs de particules) nous avons découvert successivement la suite:

\noindent {\it molécules $\rightarrow$ atomes $\rightarrow$ noyaux + électrons $\rightarrow$ protons + électrons $\rightarrow $\\ 
protons + neutrons + électrons $\rightarrow$ quarks + électrons $\rightarrow$ ??}

Nous n'avons aucune certitude que cette série se termine quelque part et, encore moins, que nous avons déjà atteint l'étape finale.  Donc nous pouvons formuler la définition suivante:

{\it Définition1: Une particule élémentaire est un objet pour lequel nous n'avons pas encore détecté de structure interne.}

Même si cette définition est correcte, elle ne décrit pas tous les objets que nous étudions dans notre Division. Une grande partie de nos recherches concernent les états hadroniques, même si nous savons parfaitement qu'ils sont en fait composés. Une définition plus “pratique”, à défaut d'être rigoureuse, serait:

{\it Définition2: Une particule élémentaire est un objet qui figure dans la “Table des Particules Elémentaires”.}

En fait, nous étudions toutes les particules qui sont “élémentaires” d'après la première définition ainsi que tous leurs états liés directs. Les états liés dont les constituants sont eux-même des états liés sont étudiés dans d'autres divisions, par ex. celles de la Physique Nucléaire ou la Physique Atomique. 

En ce qui concerne les champs, la notion même n'était pas évidente pour les pères fondateurs de la mécanique. 
Dans une lettre à Richard Bentley en 1687 Newton écrit: 

{\it “That one body may act upon another at a distance through a vacuum, without the mediation of something else, by and through which their action and force may be conveyed from one to another, is to me so great an absurdity, that I believe no man, who has in philosophical matters a competent faculty of thinking, can ever fall into it...”}

 Newton ne voulait pas admettre qu'un corps puisse créer un champ de force autour de lui et exercer ainsi sur un autre corps une action à distance, même s'il voyait la nécessité de cette notion pour sa théorie de la gravitation. Je ne sais pas à partir de quel moment l'existence des champs dynamiques fut admise indépendamment de tout support matériel, mais elle  semble être relativement récente\footnote{\`A titre d'exemple, je cite un article de Marcel Brillouin\cite{Brillouin1}  qui, en 1919, a essayé de lier les conditions de quantification de Bohr aux propriétés élastiques de l'éther.}. La mécanique quantique, qui a bouleversé plusieurs concepts de la mécanique classique, a aussi changé les notions de champs et de particules et a révélé une relation insoupçonnée entre elles. Dans cette note  je vais présenter ses évolutions en mettant l'accent sur les contributions françaises. Je vais souvent suivre, de façon approximative, l'ordre chronologique.
 
 \section{La Pré-Histoire}
 
 Même si notre Division n'existe qu'à partir de 1969, ses racines sont beaucoup plus anciennes. Dans cette note je vais  essayer de couvrir pratiquement un demi-siècle, de 1923, année de naissance de la nouvelle mécanique quantique, à 1973 et la construction théorique du Modèle Standard. 
 
 \subsection{1920 - 1930 : La mécanique quantique}\label{MQ}
 C'est probablement la décennie la plus extraordinaire, celle de la mécanique quantique et son histoire a été écrite à maintes reprises, en particulier pour la contribution française\cite{Pestre}\cite{Vila-Valls}.  En 1913 Niels Bohr avait élaboré un “modèle atomique” qui essayait de réconcilier le modèle héliocentrique de Rutherford avec la stabilité des atomes et la série de Balmer. Ce modèle, complété par A. Sommerfeld, rendait bien compte des spectres des atomes légers. Le problème était conceptuel: les règles de Bohr-Sommerfeld semblaient arbitraires, sans support théorique cohérent. A. Pais\cite{Pais} écrit pour cette période 
{\it “It was the epoch of belief, it was the epoch of incredulity”.} 

Les germes de la réponse se trouvaient déjà dans la théorie du rayonnement électromagnétique. Entre 1905 et 1909 Einstein, s'inspirant du travail de Planck,  avait démontré que les fluctuations d'énergie dans une cavité sont données par la somme de deux termes, un qui correspond à un phénomène purement ondulatoire et un deuxième qui décrirait le mouvement des quanta ponctuels\footnote{Le terme “photon” n'a été introduit qu'en 1926 par le chimiste G.N. Lewis.}. Dans son article de 1909 il conclut que “\dots la prochaine phase de la physique théorique sera une théorie de la lumière pouvant être interprétée comme une fusion entre une théorie d'ondes et une d'émission.” 

Louis de Broglie était au courant de ces travaux. Il avait aussi remarqué une analogie formelle entre l'optique physique et l'optique géométrique\footnote{Il semble que le premier à avoir fait cette remarque est W.R. Hamilton et de Broglie se réfère à lui\cite{deBroglie0}.}. La deuxième  pourrait s'obtenir de la première dans l'approximation de phase stationnaire. Dans un premier travail de 1922\cite{deBroglie1} il a retrouvé les résultats d'Einstein sous l'hypothèse que 

{\it “\dots le rayonnement noir en équilibre à la température $T$} [est] {\it un gaz formé d'atomes de lumière d'énergie $W=h\nu$.”}

 L'année suivante il franchit l'étape décisive: Dans une série de deux notes présentées par Jean Perrin à l'Académie des Sciences\cite{deBroglie2}\cite{deBroglie3}, il développe l'idée qui sera la base de l'équation de Schr\"odinger trois ans plus tard. Si à une onde électromagnétique nous pouvons associer des quantités ayant des propriétés de corpuscules, il est normal d'associer aussi une onde à une particule. Soit une particule de masse au repos $m_0$. Son énergie totale est $E_0=m_0c^2$. L. de Broglie écrit: 

{\it “\dots le principe des quantas conduit à attribuer cette énergie interne à un phénomène périodique de fréquence $h\nu_0=m_0c^2$\dots  Pour l'observateur fixe, à l'énergie totale du mobile correspondra une fréquence $\nu=\frac{m_0c^2}{h\sqrt{1-\beta^2}}$\dots 

\dots Supposons maintenant qu'au temps $t=0$ le mobile co\"{i}ncide dans l'espace avec une onde de fréquence $\nu$ ci-dessus définie se propageant dans la même direction que lui avec la vitesse $c/\beta$. Cette onde de vitesse plus grande que $c$ ne peut correspondre à un transport d'énergie; nous la considérons seulement comme une onde fictive associée au mouvement du mobile. 

Je dis que, si au temps $t=0$, il y a accord de phase entre les vecteurs de l'onde et le phénomène  interne du mobile, cet accord de phase subsistera.”}

Il va appliquer ce principe d'accord de phase d'abord aux “atomes de lumière”\footnote{Il les considère comme des particules de masse très faible, $<10^{-50}$ gr.} et il va retrouver ses résultats obtenus auparavant\cite{deBroglie1}. Ensuite aux électrons en mouvement périodique dans un atome. Il écrit:

“{\it Il est presque nécessaire} (en italique dans le texte) {\it de supposer que la trajectoire de l'électron n'est stable que si l'onde fictive \dots } [est] {\it en résonance sur la longueur de la trajectoire.”}

Cette condition lui donne les règles de Bohr-Sommerfeld. Dans la deuxième note\cite{deBroglie3} il applique le principe de phase stationnaire pour les ondes associées à des particules libres. 

{\it “En chaque point de sa trajectoire, un mobile libre suit d'un mouvement uniforme le rayon de son onde de phase, c'est à dire (dans un milieu isotrope) la normale aux surfaces d'égale phase.”}

Il prédit ainsi des phénomènes de diffraction pour des particules comme les électrons\footnote{Ce phénomène a été observé pour la première fois en 1927 par le Britanique G.P. Thomson à Aberdeen et les Américains C.J. Davisson and L.H. Germer aux laboratoires Bell.}.  Il a continué à développer ces idées et, en 1924\cite{deBroglie31}, il montre que cette loi de propagation implique l'identité des principes de Maupertuis et de Fermat.

Il ne propose pas de signification physique précise à ces ondes et cette question va le préoccuper tout le long de sa vie. Dans la note \cite{deBroglie3} il écrit:

{\it “On peut du reste considérer la vitesse $\beta c$ comme la “vitesse de groupe” d'ondes ayant des vitesses $\frac{c}{\beta}$ et des fréquences $\frac{m_0c^2}{h\sqrt{1-\beta^2}}$, correspondant à des valeurs de $\beta$ voisines mais légèrement différentes. Laissant de côté la signification physique de cette onde (ce sera la tâche difficile d'un électromagnétisme élargi de l'expliquer), nous rappelons que le mobile a même phase interne que la portion de l'onde 
située au même point; nous l'appellerons donc “l'onde de phase”.”}

Comme on pouvait s'y attendre, à cette époque “of belief and incredulity” cette idée a provoqué des réactions diverses. A Copenhague c'était plutôt la réserve. Einstein semble avoir deviné sa portée\footnote{Dans une lettre à Lorentz en 1924 il écrit “Je crois que c'est le premier faible rayon de lumière sur notre pire énigme de physique.” et à une autre à Langevin “\dots il a soulevé un coin du grand voile \dots”}, mais celui qui a saisi tout son potentiel fut Erwin Schr\"odinger. En essayant de répondre à une question de P. Debye, il a cherché l'équation d'onde qui décrirait la propagation de ces “ondes de matière”. Naturellement, au départ il a cherché une équation invariante relativiste et il s'est confronté au fait que la loi de dispersion d'une telle équation pour une particule libre de masse $m$ et d'impulsion ${\bm p}$ est $E=\pm \sqrt{{\bm p}^2+m^2}$. Schr\"odinger n'arrivait pas à donner un sens physique, ou à éliminer les solutions à énergie négative et, en 1926, a décidé de prendre la limite non-relativiste\footnote{Pour l'histoire, la version relativiste, connue aujourd'hui sous le nom d'équation de Klein-Gordon, joue un rôle très important en théorie quantique des champs.}. C'est l'équation qui porte son nom. Avec son interprétation probabiliste elle est devenue un des piliers de la mécanique quantique et a consacré la réputation de de Broglie. Ce dernier a reçu le prix Nobel de Physique en 1929. C'était le premier prix Nobel attribué à la nouvelle mécanique quantique. 

Dans cet essai nous ne pouvons pas décrire toute l'histoire de cette décennie extraordinaire et nous nous bornerons à citer quelques étapes. 

$\bullet$ En 1925 W. Heisenberg a proposé sa version de la mécanique quantique que Born et Jordan ont interprété comme une théorie de matrices. \`A première vue la mécanique de Heisenberg n'est pas une mécanique ondulatoire. Elle postule que les fonctions de la mécanique classique $\vec {x}(t)=\big (x_1(t),x_2(t),x_3(t)\big )$ et $\vec {p}(t)=\big(p_1(t),p_2(t),p_3(t)\big)$, qui décrivent, respectivement, la position et la quantité de mouvement d'une particule, deviennent des opérateurs qui satisfont à des relations de commutation
\be
\label{comrel1}
[x_a(t),p_b(t)]\equiv x_a(t)p_b(t)-p_b(t)x_a(t)=\im \hbar \delta_{\it{ab}}~~~;~~~{\it a,b}=1,2,3
\ee

$\bullet$ La même année Pauli a postulé le principe d'exclusion qui interdit à deux électrons de se trouver dans le même état quantique et Uhlenbeck et Goudsmit ont introduit la notion du spin pour un électron. Born, Heisenberg et Jordan ont développé la méthode de quantification canonique pour le rayonnement électromagnétique et établi les bases de la théorie quantique des champs.

$\bullet$  1926 est l'année de publication de l'équation de Schr\"odinger.  Fok a remarqué que si on élargit ses propriétés de symétrie à une “symétrie de jauge”, on obtient une équation décrivant le mouvement d'un électron dans un champ électromagnétique. C'était le premier exemple d'une connexion entre les propriétés géométriques et dynamiques d'une théorie.  C'est encore en 1926 que Born a proposé pour la première fois l'interprétation probabiliste de cette nouvelle mécanique et von Neumann a construit ses fondements  mathématiques. Le résultat le plus marquant de ce travail est qu'en mécanique quantique un état physique est décrit par un vecteur dans un espace de Hilbert.

$\bullet$ La même année Léon Brillouin a publié aux CRAS\cite{Bril1} un article dans lequel il montre que {\it “\dots l'équation} [de Schr\"odinger] {\it peut être résolue par approximations successives, la première approximation redonnant l'ancienne mécanique quantique.”}  La même méthode a été proposée indépendamment par H.A. Kramers et G. Wentzel et elle connue comme “approximation BKW”.

$\bullet$ L'année suivante Heisenberg a obtenu les célèbres relations d'incertitude et Bohr a formulé le principe de complémentarité. 

$\bullet$  En 1927 de Broglie revient sur la mécanique quantique après la publication de l'équation de Schr\"odinger. Dans un article publié dans le Journal de Physique\cite{deBr32}, il conteste l'interprétation donnée à la fonction d'onde et écrit:

{\it “Le but du présent mémoire est de montrer que les solutions continues fournissent, en réalité, seulement une certaine vue statistique des phénomènes dynamiques dont la description exacte exige sans doute la considération d'ondes comportant des singularités.”} 

C'est le début de la théorie connue sous le nom de {\it “double solution”} ou {\it “théorie de l'onde pilote”}. L. de Broglie essaye de concilier la propagation d'une particule décrite par  l'équation d'onde de la mécanique quantique avec l'idée intuitive d'une “trajectoire” associée au mouvement d'une particule massive en mécanique classique. Il a présenté cette idée au Congrès Solvay de 1927\cite{Solvay27} comme une alternative au principe de complémentarité, sans arriver à convaincre les participants. Néanmoins, nous allons voir plus tard que cette idée, ressuscitée dans les années 50, a motivé des travaux qui ont ouvert un nouveau champ de recherches dans lequel des chercheurs français ont joué un rôle de pionnier. 

$\bullet$ Entre 1927 et 1928 Dirac a entrepris la recherche d'une équation d'onde invariante relativiste, mais avec un avantage décisif: on savait déjà que l'électron avait un spin égal à 1/2 et, par conséquent, sa fonction d'onde avait deux composantes, une pour décrire le mouvement de l'électron avec projection de spin +1/2 et une autre pour la projection -1/2. Donc Dirac a cherché une équation pour une fonction multi-composante. Contrairement à Schr\"odinger, Dirac eut l'idée de chercher une équation du premier ordre. Il est facile à voir qu'une telle équation pour une fonction à une composante serait triviale, mais Dirac a démontré qu'on peut obtenir une équation intéressante à condition que la fonction d'onde soit un objet à quatre composantes. C'est l'équation qui porte son nom. En faisant cela, Dirac a découvert les représentations spinorielles du groupe de Lorentz. Avec son équation il a obtenu d'emblée deux résultats confirmés par l'expérience: il a donné la valeur correcte (avec la précision de l'époque) du moment magnétique de l'électron\footnote{Cette prédiction de la valeur du moment magnétique $g=2$ n'est pas spécifique d'une théorie relativiste, mais ceci a été découvert beaucoup plus tard\cite{Lev-Lebl}.} et, surtout, il a prédit l'existence d'un “anti-électron”. La découverte en 1932 du positron par C. Anderson lui a donné la consécration suprême. 

C'est encore en 1928 que Dirac utilisa la théorie quantique du champ électromagnétique, proposée par Born, Heisenberg et Jordan en 1925, pour décrire l'émission spontanée d'un photon par un atome qui se trouve dans un état excité. C'était la première application de la théorie quantique des champs. Avec ce calcul Dirac a confirmé la correspondance entre le champ électromagnétique et les photons et il a montré le mécanisme qui permet la création, et l'absorption, de ces derniers.

$\bullet$ En parallèle, durant la même décennie ont été développées les statistiques quantiques sous forme des règles de comptage pour les états d'un ensemble de particules identiques. Par Bose et Einstein en 1924 pour les quanta de lumière et par Fermi en 1926 pour les électrons. La même année,  Dirac a interprété ces règles comme propriétés de symétrie (ou anti-symétrie) de la fonction d'onde et, en 1928, Jordan et Wigner ont introduit le formalisme des anti-commutateurs. Il faut remarquer que la connexion entre spin et statistique n'était pas faite: les règles de Bose-Einstein étaient sensées décrire un ensemble de photons et celles de Fermi-Dirac un ensemble de particules. Lorsque, en 1928, Dirac a voulu décrire la statistique d'un gaz d'He$^4$ qui a spin égal à 0, il a utilisé la statistique de Fermi-Dirac. Une première formulation du théorème qui relie spin et statistique est due à Pauli et date de 1940. 

$\bullet$ En 1930 L. Brillouin étudia la propagation d'électrons dans un cristal et établit l'existence des “zones de Brillouin”\cite{Brill}, un concept fondamental en physique du solide.

\subsection{1930 - 1940 : La physique nucléaire}

C'est la décennie de la consolidation de la mécanique quantique et de la naissance de la physique nucléaire. Des nouveaux concepts ont enrichi la physique fondamentale et les premiers cyclotrons ont été construits. Nous ne présenterons pas toutes ces découvertes en détail parce que les contributions des chercheurs français n'ont pas été parmi les plus significatives.

\subsubsection{Le neutrino}  L'étude du rayonnement $\beta$ a été une source inépuisable de découvertes.  
Aujourd'hui nous savons qu'il s'agit de la désintégration du neutron de la forme: $n\rightarrow p~+~e^-~+~\bar{\nu}$. La même réaction, pour un neutron lié à un noyau, apparaît comme $N_1\rightarrow N_2~+~e^-~+~\bar{\nu}$.
Pour les physiciens des trois premières décennies du XXe siècle, cette réaction n'a cessé de leur poser problème.  (i) Ils ne connaissaient ni le neutron ni le neutrino et, en plus, ce dernier, avec les moyens de l'époque, était indétectable. Donc ils voyaient uniquement: $N_1\rightarrow N_2~+~e^-$.  (ii) Comme ils ne pouvaient pas imaginer qu'un électron puisse sortir d'un noyau sans y être déjà dedans,  ils avaient un modèle nucléaire selon lequel les noyaux étaient composés de protons et d'électrons. Mais le problème le plus sérieux était que, d'après des mesures d'une grande précision, cette désintégration semblait violer la conservation de l'énergie et du moment cinétique. Plusieurs théoriciens, dont Niels Bohr, pensaient que toutes ces lois de conservation n'étaient plus valables en mécanique quantique. La solution fut trouvée par Pauli en Décembre 1930. Il proposa l'existence d'une troisième particule, baptisée “neutrino” (petit neutron en italien) par Fermi, qui était émise avec l'électron mais qui échappait à la détection\footnote{Il a fallut attendre 1956 et le développement des premiers réacteurs nucléaires pour que la détection des neutrinos devienne possible.} et emportait l'énergie manquante. Pauli pensait que les neutrinos, tout comme les électrons, faisaient partie de la matière nucléaire. L'existence d'une telle particule fut rejetée par plusieurs grands physiciens, dont Niels Bohr, qui continua à proposer des théories violant la conservation de l'énergie jusqu'en 1938\footnote{A. Pais\cite{Pais} écrit: {\it it is clear that particles and fields belong to the post-Bohr era.}}.

De cette période nous pouvons citer deux travaux des théoriciens français. 

1) En 1933 Francis Perrin, dans une note à l'Académie des Sciences\cite{Perrin1} présentée par Jean Perrin, propose pour la première fois que les neutrinos soient créés au moment de l'émission: 

{\it “Si le neutrino a une masse intrinsèque nulle on doit aussi penser qu'il ne préexiste pas dans les noyaux atomiques, et qu'il est créé, comme l'est un photon, lors de l'émission. Enfin, il semble qu'on doive lui attribuer un spin 1/2\dots” }

Dans cette note Perrin remet en question la croyance de l'époque, selon laquelle tout ce qui sort d'un noyau se trouve déjà dedans, mais il limite cette possibilité de création aux particules de masse nulle. 

2) En 1934 L. de Broglie\cite{deBroglie4} présenta un modèle du photon comme état lié neutrino--anti-neutrino. Dans cet article le terme “anti-particule” est utilisé pour la première fois.  

On aura souvent l'occasion de parler de neutrinos dans la suite.

\subsubsection{ Les symétries en physique nucléaire}  1932 est l'année de la découverte du neutron. Ses propriétés étaient mal connues et beaucoup de physiciens le considéraient comme un état lié proton-électron, une sorte de minuscule atome d'hydrogène. Heisenberg partageait cette opinion. Dans la controverse qui opposait Bohr et Pauli à propos du neutrino, il se rangeait du côté de son maître Bohr. Il écrit : {\it “\dots sous des conditions adéquates, le neutron peut se désintégrer en un proton et un électron et, dans ce cas, les lois de conservation de l'énergie et du moment cinétique ne s'appliquent probablement pas.”} Néanmoins, en 1932 il introduisit un nouveau concept de symétrie qui s'est révélé être extrêmement fécond.  Il postula que le système proton-neutron se comporte comme s'il y
avait une seule particule, qu'on appellera ici {\it nucléon}, et qui peut se
présenter en deux états, soit comme {\it proton}, soit comme {\it
neutron}. Dans ce travail Heisenberg a introduit pour la première fois une quantité physique qui n'est pas du tout connectée à l'espace ordinaire. Nous l'appelons aujourd'hui  {\it 
spin isotopique}, et nous lui attribuons, formellement, les mêmes propriétés que le
spin ordinaire.  Tout comme le spin, le spin isotopique devrait être une quantité vectorielle,
mais, et c'est ici le point essentiel, il devrait être un vecteur non pas de l'espace
ordinaire, mais d'un autre espace, abstrait, {\it l'espace
isotopique}. C'est l'idée d'une {\it symétrie interne}. Pour la première fois en physique nous avons
considéré des changements d'un système des coordonnées autres que celui de
notre espace-temps ordinaire. Dans l'article de Heisenberg la symétrie n'était pas respectée; il y avait des forces entre un proton et un neutron, ainsi qu'entre deux neutrons, mais pas entre deux protons. Nous verrons dans la suite l'évolution de cette idée fondamentale. 

\subsubsection{Fermi et la théorie des interactions faibles}
Déjà en 1926, avant l'équation de Schr\"odinger, Fermi publia deux articles contenant les règles statistiques pour un ensemble d'électrons. C'est l'origine du terme {\it fermions.} En 1933 il est allé plus loin avec un article parmi les plus influents de la physique des particules dans lequel il propose un modèle de théorie des champs pour décrire la désintégration $\beta$ du neutron. Même aujourd'hui, après la construction du Modèle  Standard, la théorie de Fermi est utilisée comme une bonne approximation à basse énergie.

Le papier, sous le titre {\it Tentativo di una teoria della emissione di raggi $\beta$} \footnote{Une version anglaise avait été soumise à {\it Nature} mais elle a été rejetée {\it “\dots because it contained speculations too remote from reality to be of interest to the reader.”}}, contient   plusieurs idées révolutionnaires. Fermi était un des premiers physiciens à accepter l'existence physique du neutrino. Dans cet article il rompt complètement avec l'idée selon laquelle toutes les particules qui sortent d'un noyau doivent nécessairement se trouver à l'intérieur. Il développe la théorie quantique des champs pour les fermions et introduit le formalisme des opérateurs de création et d'annihilation. C'est l'article fondateur qui a établi la théorie quantique des champs en tant que langage universel de la physique fondamentale.

\subsubsection{Yukawa et les forces nucléaires} En physique classique les forces entre deux corps chargés sont décrites par le champ électromagnétique. En physique quantique on fait correspondre au champ une particule, le photon, et on dit que les deux corps échangent un, ou plusieurs photons. On peut démontrer que le long rayon d'action des forces électromagnétiques est dû à la masse nulle du photon. En 1935 Hideki Yukawa a généralisé ce concept aux forces nucléaires. Expérimentalement on savait qu'elles sont de courte portée, de l'ordre de 10$^{-15}$ m, donc il a postulé l'existence d'une particule, appelée “méson”, de masse de l'ordre de 100-200 MeV, et dont l'échange entre protons et neutrons produirait les forces nucléaires. 

Dans son premier article Yukawa pensait que ce méson, tout comme le photon, avait spin égal à 1, mais plus tard ceci a été corrigé à un spin égal à 0 \footnote{Pour l'histoire, en 1936 une nouvelle particule découverte avec le rayonnement cosmique fut interprétée comme “le méson de Yukawa”. Cependant, après la guerre il s'est avéré qu'elle était en fait un nouveau lepton, comme l'électron, avec spin égal à 1/2. Le vrai méson, appelé “méson $\pi$”,  fut découvert lui aussi avec le rayonnement cosmique en 1947. Il a effectivement spin égal à 0. C'est la dernière particule découverte avec les rayons cosmiques. Depuis, les accélérateurs ont pris la relève.}.

De cette période nous voulons citer l'article de A. Proca\cite{Proca1}, physicien d'origine Roumaine travaillant à l'IHP. Il avait soutenu sa thèse en 1933 sous la direction de de Broglie et, en 1936, publia l'équation qui porte son nom et décrit le mouvement des particules massives de spin égal à 1. Même si à l'origine Proca pensait que cette équation pourrait s'appliquer aux électrons et aux positrons, c'est l'équation qui est utilisée aujourd'hui pour les bosons intermédiaires $W^{\pm}$ et $Z^0$.  

\subsubsection{Kemmer et la symétrie d'isospin des forces nucléaires} 
Entre 1932 et 1938 trois développements importants ont permis à l'idée originale de Heisenberg de devenir l'invariance par isospin des forces nucléaires qu'on connaît aujourd'hui. Un laps de temps extrêmement court, vu le caractère révolutionnaire du concept.

Le premier, et probablement le plus important, est le progrès concernant les techniques expérimentales qui ont fourni une vaste variété de données. Elles ont montré la nécessité d'une force proton-proton et l'indépendance de charge des forces nucléaires. En 1936 ce résultat était fermement établi. Dans un article de B. Cassen et E. U. Condon on trouve la formulation complète de l'indépendance de charge et l'extension du principe d'exclusion de Pauli aux variables d'isospin. Le deuxième est la théorie de Fermi qui a fourni le cadre de la théorie quantique des champs. Le troisième est l'hypothèse de Yukawa sur l'existence des mésons. 

La synthèse de tous ces ingrédients fut l'oeuvre de Nicholas Kemmer en 1938. Il a eu l'idée d'incorporer les mésons de Yukawa au formalisme de Heisenberg et d'écrire la première interaction pion-nucléon invariante par isospin. Il a utilisé la théorie quantique des champs en introduisant des champs quantiques pour les nucléons et les pions. A cette époque il y avait une évidence expérimentale, même si elle était erronée, pour l'existence des mésons chargés. Kemmer a compris que l'invariance par isospin nécessitait l'introduction aussi d'un méson neutre $\pi^0$ qui a été découvert en 1950 au synchrotron à électrons de Berkeley par Panofsky, Steinberger et Steller.  
Le $\pi^0$ devint ainsi la première particule dont l'existence fut prédite par un argument de symétrie et la première à être découverte auprès d'un accélérateur\footnote{Kemmer n'a pas reçu la reconnaissance que son travail mérite. Ses équations sur les forces pion-nucléon se trouvent dans tous les livres, mais son nom n'est presque jamais mentionné.}.

L'espace isotopique est tri-dimensionnel, isomorphe à l'espace ordinaire. Mais au fil du temps, avec la découverte des nouvelles symétries internes, l'espace dans lequel agissent les transformations physiques est devenu une variété mathématique multi-dimensionnelle, avec des propriétés topologiques variées, dont une partie seulement, l'espace-temps à quatre dimensions, est directement accessible à nos sens. 

\subsubsection{Wigner et les représentations du groupe de Poincaré}
En 1939 E. P. Wigner publia la première étude complète des représentations linéaires du groupe de Poincaré. Malgré sa grande importance pour la physique, ce chapitre des mathématiques appliquées était mal connu et on avait uniquement des résultats partiels qui prêtaient souvent à  confusion.

\subsection{1940 - 1950 : La naissance d'une discipline}\label{40-50}
Sous sa forme moderne, la physique théorique des interactions fondamentales
a une date de naissance très précise: le 2 juin 1947, la 
conférence de Shelter Island. C'était la première conférence scientifique organisée après la guerre par l' Académie des Sciences des Etats Unis avec le titre “Les fondements de la mécanique quantique”\footnote{La conférence de Shelter Island était sensée initier une série sous le même titre. En effet elle a été suivie en 1948 par une deuxième à Pocono (Pennsylvanie) et en 1949 par une troisième à Oldstone (NY). Après le sujet des fondements de la mécanique quantique avait été déclaré clos. Vision très optimiste, bien sûr!}. Elle a réuni autour de R. Oppenheimer 24 physiciens, venant tous des USA, du 2 au 4 juin 1947. Les contributions les plus importantes annoncées
à cette conférence n'étaient pas de grandes avancées théoriques, mais deux
résultats expérimentaux. L'existence  d'un déplacement entre les niveaux
$2^2S_{1/2}$ et $2^2P_{1/2}$ de l'atome d'hydrogène, qu'on appelle “le Lamb shift” et  la première mesure du moment magnétique “anormal” de l'électron.  L'intérêt de ces résultats était dû au fait que, pour la première
fois, ils étaient en contradiction flagrante avec la théorie de Dirac qui était “la
Bible” de la physique théorique de l'époque. Ils ont forcé les 
théoriciens à faire face aux problèmes des divergences de la théorie quantique des champs et à élaborer une théorie cohérente qui a
révolutionné nos idées sur toute la physique fondamentale. C'est la théorie de la “renormalisation” qui fut développée par Feynman, Schwinger et  Tomonaga en 1947 et complétée une année plus tard par Dyson.

La formulation générale de la théorie quantique des champs était connue depuis le travail de Fermi en 1933 mais son application pratique butait sur des problèmes techniques dont l'origine est facile à comprendre.

Les deux potentiels classiques, l'électrostatique et le gravitationnel, sont de la forme $V(r)\sim 1/r$. Les deux sont singuliers lorsque $r\rightarrow 0$. C'est la cause de l'instabilité de l'atome de Rutherford. La mécanique quantique résout ce problème. En résolvant l'équation de Schr\"odinger on trouve un état fondamental à énergie finie. En fait, la mécanique quantique, en introduisant une géométrie non-commutative dans l'espace des phases, conduit à une interprétation des observables physiques en tant qu'opérateurs dans un espace de Hilbert. Ces opérateurs peuvent avoir une représentation matricielle et un spectre discret de valeurs propres. On s'est vite aperçu que cette solution est partielle; elle se limite à certains potentiels et elle n'est valable qu'à l'approximation statique. Au delà de cette approximation des nouvelles divergences apparaissent  qui dépendent de la théorie et qui ont hanté la théorie quantique des champs pendant longtemps. 

Lorsque dans le calcul d'une quantité physique on trouve un résultat divergent, c'est parce qu'on a fait une erreur mathématique. Ici l'erreur est évidente: c'est l'approximation de la particule “ponctuelle” qui conduit à une densité d'énergie et de charge infinie. La théorie de la renormalisation consiste à donner une définition mathématiquement correcte des expressions qui apparaissent dans les calculs de la  théorie quantique des champs et, du point de vue physique, à préciser quelles sont les quantités qui sont mesurables. C'est un processus technique assez complexe qui ne peut s'appliquer qu'à un très petit nombre de théories\footnote{Ce nombre dépend de la façon de compter les théories, mais, en général, il est considéré égal à 6.}. L'électrodynamique quantique, c'est à dire la théorie qui décrit l'interaction entre les électrons et le champ électromagnétique, en fait partie et elle a fourni la première confirmation éclatante de cette technique. Par exemple la valeur expérimentale du moment magnétique “anormal” (c.-à-d. la différence avec la valeur donnée par l'équation de Dirac) de l'électron est $(g-2)/2=(1159.65218091 ± 0.00000026) × 10^{-6}$, à comparer avec la prédiction théorique qui est $(g-2)/2=(1159.652181643 ± 0.000000764) × 10^{-6}$. Un autre fait remarquable est que la Nature utilise toutes les théories renormalisables et elles seules, pour décrire les interactions entre les particules élémentaires; avec toutefois une exception notable: la théorie de la gravitation. En effet, la version quantique de la relativité générale n'est pas une des six théories renormalisables et, par conséquent, nous ne connaissons pas de théorie quantique du champ gravitationnel. 
\vskip 0.3cm
Avant de quitter cette section je voudrais mentionner une autre grande contribution de Feynman à la théorie quantique. En 1948, développant une idée de Dirac, il a proposé une nouvelle règle de quantification, indépendante de la quantification canonique de Heisenberg. Elle s'appelle “quantification par l'intégrale de chemins” et elle s'inspire du phénomène de diffraction pour les électrons prédit par de Broglie. Les deux méthodes sont équivalentes dans tous les cas où les deux s'appliquent, mais la méthode de Feynman présente certains avantages: (i) elle est plus facile à appliquer à la quantification des théories avec contraintes, par ex. les théories de jauge et, (ii)  elle n'est pas liée à la théorie des perturbations et se prête mieux aux simulations numériques dans le régime du couplage fort.

\section{Les Temps Modernes}
Pour la physique des particules élémentaires l'année 1950 marque le passage de la “préhistoire” aux “temps modernes”. C'est l'année de la découverte du pion neutre, la première particule qui a été découverte à l'aide d'un accélérateur  (le pion chargé avait été découvert avec les rayons cosmique en 1947\footnote{Le premier cyclotron était construit à Berkeley en 1930 par E. O. Lawrence et son étudiant M. S. Livingstone. Entre 1930 et 1940 Laurence avait construit dans son laboratoire une série de cyclotrons de plus en plus puissants qui étaient capables de produire abondamment des pions, mais les physiciens n'avaient pas encore développé des techniques de détection appropriées. Il ne suffit pas d'avoir un nouveau jouet, il faut encore apprendre à jouer avec. Il est vrai aussi que Lawrence s'intéressait surtout à la construction de machines et pas tellement aux expériences qu'on pouvait faire avec elles.}). Une des conséquences de l'arrivée en puissance des accélérateurs a été la séparation des deux communautés, celle des particules et celle des rayons cosmiques\footnote{La dernière grande conférence qui a réuni les deux communautés eut lieu en France, à Bagn\`eres-de-Bigorre en Juillet 1953. Elle était organisée par L. Leprince-Ringuet, et des physiciens des rayons cosmiques lui avaient reproché d'avoir accordé trop de temps aux résultats, encore maigres, venant des accélérateurs. C.F. Powell, à l'origine de la découverte du méson $\pi^+$ et un grand nom de la recherche avec les rayons cosmiques, déclara dans son discours de clôture:  “Gentlemen, we have been invaded\dots The accelerators are here.” Une mesure de cette “invasion”: à la même Conférence l'Australien R.H. Dalitz présenta une méthode astucieuse pour l'étude  des nouvelles particules qui venaient d'être découvertes avec les rayons cosmiques et qui se désintégraient en trois corps. C'est le “Dalitz plot” dans lequel chaque événement  est représenté par un point sur un graphique. En 1953 Dalitz avait 13 points, tous venant des rayons cosmiques. A la Conférence de Rochester de 1955 il en avait 53 avec 42 venant encore des cosmiques. L'année suivante il en avait plus de 600, dont la grande majorité venaient des accélérateurs. Il faut néanmoins noter que ces deux communautés se sont en grande partie retrouvées récemment avec l'émergence d'une nouvelle discipline, “les Astroparticules”.}. Ce qui est plus important est que la pratique de la physique expérimentale fut profondément affectée. Avant les accélérateurs une équipe aurait conçu et réalisé une expérience, ainsi qu'analysé les résultats, dans leur laboratoire. Avec les accélérateurs ces fonctions ont été en grande partie dissociées. Des grands Centres de Recherche, nationaux ou même internationaux, ont été créés -- Brookhaven aux USA, CERN en Europe -- qui n'étaient pas associés à une université. La complexité des systèmes de détection et la cadence avec laquelle les données arrivaient, ont conduit à la création des grandes collaborations internationales. Nous sommes entrés dans l'ère de la “Big Science”. La conséquence la plus frappante fut la découverte d'un grand nombre des nouvelles particules “élémentaires”. En 1940 on connaissait le proton, le neutron, l'électron, le neutrino (pas encore détecté) et la particule sensée être le méson de Yukawa. Dans les années 40 certaines “nouvelles particules”, pas toujours bien identifiées, ont été signalées grâce aux rayons cosmiques. Ce sont les particules qu'on appelle aujourd'hui “particules étranges”\footnote{Il semble que la première observation a été faite en France\cite{LePrince} pendant la guerre.}.  Avec les accélérateurs, les découvertes se sont succédées à un rythme exponentiel: en 1960 la “Table des Particules Elémentaires” comptait des dizaines d'éléments et aujourd'hui des centaines, même si nous savons  que tous ces états ne sont point “élémentaires”. 
\vskip 0.3cm
En France cette période a marqué l'internationalisation de la physique des particules avec deux points essentiels: l'adoption de l'anglais et l'abandon des journaux nationaux pour toute publication jugée “importante”. En étudiant les CRAS ou Le Journal de Physique nous constatons une baisse du nombre et du niveau des articles publiés en physique des hautes énergies. Vers la fin des années  50, cette discipline avait pratiquement disparu des publications françaises. La physique théorique a suivi ce mouvement avec, principalement, des acteurs qui avaient effectué des séjours de formation à l'étranger. Nous allons présenter ces évolutions en essayant toujours de souligner 
les contributions françaises. 

\subsection{1950 - 1960 : Les accélérateurs - Création du CERN}
C'est la décennie pendant laquelle la physique des particules élémentaires a atteint l'âge adulte, aussi bien sur le plan expérimental que sur celui de la théorie, avec une collaboration très étroite entre les deux. Nous commençons par un panorama de la situation internationale et nous finirons par ce que nous appellerons “le renouveau français”.

\subsubsection{La situation internationale}
\label{50-60-Int}

Nous ne pouvons pas présenter ici toutes les découvertes de cette grande décennie et nous nous limiterons à une liste parmi les plus significatives.
\vskip 0.3cm
$\bullet$ {\bf Les particules étranges.} Parmi les nouvelles particules découvertes d'abord avec les rayons cosmiques et étudiées surtout auprès des  accélérateurs, certaines présentaient un comportement étrange, elles étaient abondamment produites lors des collisions hadroniques, signe d'interactions fortes, mais avaient des longues vies moyennes, caractéristiques d'interactions faibles. L'explication fut donnée par M. Gell-Mann et K. Nishijima. Ils ont postulé l'existence d'un nouveau nombre quantique, appelé “étrangeté” par Gell-Mann, conservé par les interactions fortes et électromagnétiques mais violé par les interactions faibles. Gell-Mann, A. Pais et O. Piccioni ont décrit le premier phénomène d'oscillations quantiques entre particules élémentaires, à savoir entre un méson étrange neutre $K^0$ et son antiparticule $\bar{K^0}$. Ces découvertes créèrent un nouveau domaine de recherche, celui “des saveurs lourdes”. La première particule qui appartient à cette classe est le muon, découvert en 1936 et mal interprété comme étant le méson de Yukawa. C'est la première particule dont le rôle dans la structure de la matière est toujours inconnu\footnote{Voir la question célèbre d'I. Rabi : “Who ordered that?”}. La découverte des particules étranges, ainsi que d'autres encore plus lourdes, n'a pas élucidé cette question.
\vskip 0.3cm
$\bullet$ {\bf Les théories de jauge non-abéliennes - théories de Yang-Mills.} L'invariance de l'électrodynamique par des transformations locales était connue déjà au 19ème siècle. Le même type d'invariance est présent aussi dans la théorie classique de la relativité générale. Nous avons vu qu'en 1926 V. Fok avait remarqué que le fait d'imposer une invariance de phase locale à la fonction d'onde de l'équation de Schr\"odinger est équivalent à introduire un champ électromagnétique. L'extension de cette approche à l'invariance par isospin est le travail de C. N. Yang et R. Mills en 1954\footnote{Il y a eu au moins deux tentatives antérieures, une par O. Klein en 1937 et une autre par W. Pauli en 1953, mais les deux ont suivi une méthode compliquée et contre-intuitive qui consistait à passer par la relativité générale dans un espace à cinq ou six dimensions, compactifier les dimensions supplémentaires et prendre la limite de courbure nulle.}. L'importance de cet article n'a pas besoin d'être soulignée. 
\vskip 0.3cm
$\bullet$ {\bf Le neutrino.} En 1956 F. Reines et C.L. Cowan, travaillant auprès du réacteur nucléaire de Savannah River, observèrent pour la première fois une réaction causée par un neutrino. L'année suivante une collaboration menée par M. Goldhaber a réussi l'exploit de mesurer l'hélicité du  neutrino émis lors  d'une désintégration $\beta$. En 1957 B. Pontecorvo suggéra la possibilité d'oscillations pour les neutrinos. La suggestion initiale concernait des oscillations $\nu \leftrightarrow \bar{\nu}$, mais plus tard, il a étendu cette suggestion aux oscillations entre espèces de neutrinos différents. 
\vskip 0.3cm
$\bullet$ {\bf La violation de la parité et la théorie des interactions faibles.}
La découverte la plus surprenante de toute cette période fut celle de la violation de la symétrie par inversion des coordonnées de l'espace, la “parité” $P$. En essayant de comprendre certains aspects bizarres des désintégrations des mésons $K$ chargés, T.D. Lee et C.N. Yang ont remarqué qu'aucun résultat expérimental ne permettait d'affirmer que les interactions faibles respectaient l'invariance par parité. Ils ont proposé des expériences pour répondre à la question. C'est C.S. Wu, en mesurant la distribution angulaire des électrons émis lors d'une désintégration $\beta$ de cobalt polarisé, qui a montré que $P$ était bel et bien violée. Ce résultat a eu l'effet d'une bombe\footnote{Il paraît que Pauli, en apprenant ce résultat, s'est mis à crier que c'était de la foutaise!}, mais il a été confirmé par d'autres mesures indépendantes. Entre 1956 et 1957 S.S. Gershtein et Y.B. Zeldovich en l'URSS, ainsi que Feynman et Gell-Mann aux USA, ont formulé un modèle théorique pour les interactions faibles en termes d'interaction entre courants, analogues au courant électromagnétique. La violation de la parité fait que ces courants sont formés par une superposition des courants vectoriels et axiaux. C'est la théorie $V-A$. 
Ces courants, le vectoriel et l'axial, ont une partie leptonique et une autre hadronique. Il est facile d'écrire la première en termes des champs des leptons connus, l'électron, le muon, le tau, ainsi que leurs neutrinos. Nous pouvons supposer que ces champs satisfont à des équations de champs libres, ce qui nous permet de calculer les éléments de matrice de ces courants entre des états leptoniques. En revanche, la partie hadronique pose problème parce que nous ne pouvons pas supposer d'équations des champs libres pour les hadrons. Dans les années cinquante la moitié de ce problème fut résolu en faisant appel à un argument de symétrie. En comparant des mesures de la désintégration du muon -- un processus purement leptonique -- avec celles de la désintégration $\beta$, on pouvait conclure que la partie vectorielle du courant hadronique était un courant conservé. C'est la théorie CVC (Courant Vectoriel Conservé). En règle générale, d'après le théorème de Noether, la conservation d'un courant est la conséquence d'une symétrie. Si nous prenons l'exemple de la désintégration $\beta$, le courant hadronique faible transforme un neutron à un proton, par conséquent la symétrie sous jacente est celle d'isospin. Donc l'hypothèse du CVC a permis d'identifier le courant hadronique vectoriel des interactions faibles avec une partie du courant d'isospin. La solution de l'autre moitié du problème, à savoir celui de la partie axiale du courant, fut trouvée la décennie suivante.

\vskip 0.3cm
$\bullet$ {\bf Le groupe de renormalisation.} Une théorie des champs dépend d'un certain nombre de paramètres dont la valeur est déterminée par l'expérience. Ce sont d'habitude les masses de particules et les constantes de couplage qui traduisent l'intensité de l'interaction. En 1953 deux chercheurs Suisses, E. C. G. St\"uckelberg et A. Petermann, ont remarqué qu'au niveau quantique le processus de renormalisation que nous avons présenté à la section \ref{40-50}, introduit un certain arbitraire dans la définition de ces constantes. Ils ont montré que nous pouvons représenter cet arbitraire par le choix d'un paramètre non-physique qui a les dimensions d'une masse et qu'on appelle “point de soustraction”.  On obtient ainsi une famille de théories dépendant du choix de ce point et elles sont toutes liées par des transformations qui forment un groupe. C'est “le groupe de renormalisation”. Ce concept semblait avoir une signification purement formelle mais, l'année suivante, Gell-Mann et F. Low ont  montré que, sous certaines conditions, on peut interpréter les transformations du groupe de renormalisation comme des changements d'échelle d'énergie. C'est le résultat qui a fait de ce concept a priori technique, un outil extrêmement puissant pour l'étude des phénomènes très variés, du comportement des amplitudes de diffusion à haute énergie, aux transitions de phase de la matière condensée. Nous aurons l'occasion d'aborder certaines de ces applications dans la suite de cet exposé. 
\vskip 0.3cm
$\bullet$ {\bf Les bases mathématiques de la théorie quantique des champs.}  Vers 1955 A. S. Wightman proposa une formulation axiomatique de la théorie quantique des champs sous forme d'un ensemble de six axiomes. Ils sont basés sur des hypothèses physiques simples comme l'invariance par des transformations de Poincaré, la positivité de l'énergie, la propriété de localité des interactions - c'est à dire l'absence de corrélation entre des événements séparés par des séparations du genre espace - etc. En fait, ces axiomes semblent avoir un contenu tellement général qu'on ne s'attendrait pas à obtenir d'eux de résultats intéressants. Or, c'est le contraire qui est vrai. Nous pouvons démontrer deux théorèmes, avec des conséquences physiques importantes. Le premier s'appelle {\it théorème CPT} \footnote{$P$ est l'opération parité $\vec{x}\rightarrow -\vec{x}$, $T$ est l'opération qui renverse le sens du temps $t \rightarrow -t$ et $C$ est une opération qui échange une particule avec son anti-particule.} et assure que toute théorie des champs qui satisfait aux axiomes de Wightman est automatiquement invariante sous le produit de ces trois opérations prises dans un ordre quelconque. Le deuxième est le théorème {\it spin -- statistique} qui nous dit que les particules de spin entier satisfont à la statistique de Bose-Einstein et celles de spin demi-entier à celle de Fermi-Dirac. Les deux ont des conséquences physiques importantes comme, par exemple, l'égalité des masses entre particules et anti-particules pour le théorème $CPT$ et les phénomènes de condensation de Bose-Einstein pour la connexion entre spin et statistique. 

La grande majorité des résultats expérimentaux en physique des particules sont obtenus par des expériences de diffusion. Les fondements mathématiques de la théorie de la diffusion dans le cadre de la théorie quantique des champs sont obtenus dans une série de travaux initiés pendant cette décennie par R. Haag et D. Ruelle ainsi que H. Lehmann, K. Symanzik et W. Zimmermann. La théorie LSZ se trouve dans tous les livres de  théorie des champs.
\vskip 0.3cm
$\bullet$ {\bf La théorie de la matrice $S$.} Après le succès de l'électrodynamique quantique il était normal d'essayer d'appliquer la méthode de la renormalisation aux autres interactions, à savoir l'interaction forte et l'interaction faible. A l'époque l'interaction forte était représentée par celle entre les nucléons (c.-à-d. le proton et le neutron) et les mésons $\pi$. C'est une de six théories des champs renormalisables, donc, techniquement, le programme s'applique. Le problème est numérique. Nous ne connaissons pas de solutions exactes pour les théories quantiques des champs qui nous intéressent et nous faisons appel à {\it la théorie des perturbations} qui consiste à calculer les premiers termes du développement en puissances d'un paramètre qu'on appelle {\it constante de couplage.} C'est le paramètre qui détermine la force de l'interaction. Pour l'électrodynamique ce paramètre est $\alpha =e^2/4\pi \sim 1/137$, donc on peut avoir des résultats fiables même en calculant un petit nombre de termes. Pour l'interaction pion-nucléon le paramètre correspondant est de l'ordre de 10 et le développement perturbatif n'a aucun sens. Bien sûr, c'est la raison pour laquelle les interactions fortes sont fortes! Restaient les interactions faibles qui, comme leur nom l'indique, sont beaucoup plus faibles. Malheureusement, ici c'est la méthode qui ne s'applique pas. La théorie de Fermi, qui décrivait bien la phénoménologie des interactions faibles, n'est pas parmi les six théories renormalisables. Ce double échec a vite terni l'enthousiasme soulevé par la théorie de la renormalisation. Le désenchantement était 
tel que la théorie quantique des champs n'était même pas enseignée dans de nombreuses universités.  

Une approche alternative consiste à éviter la théorie quantique des champs et à chercher des méthodes de calcul applicables directement sur les amplitudes de diffusion. C'est la théorie de la matrice $S$, $S$ pour “scattering”. Elle a été introduite par J. A. Wheeler dans les années 30 et plus concrètement par Heisenberg en 1943, mais elle a été développée durant les années 50. Sans faire appel à des modèles dynamiques particuliers, elle postule directement des propriétés d'analyticité pour les amplitudes de diffusion et permet ainsi de les exprimer à l'aide des relations de dispersion qui étaient connues en optique. Par exemple, on peut démontrer que les propriétés d'analyticité qui permettent d'écrire les relations de dispersion de Kramers-Kronig décrivant la diffusion de la lumière dans la matière, peuvent être obtenues à partir de la propriété de causalité\footnote{Cette observation est probablement due à van Kampen qui était un étudiant de Kramers.}. 

La première relation de dispersion des temps modernes fut écrite par M. Goldberger en 1955 pour l'amplitude de diffusion pion-nucléon vers l'avant. En 1958 S. Mandelstam a généralisé ce travail en postulant des propriétés d'analyticité de l'amplitude de diffusion considérée comme fonction de deux variables complexes, l'énergie et le moment de transfert. Cette hypothèse lui permit d'écrire une relation de dispersion double, connue sous le nom de “représentation de Mandelstam”. L'année suivante T. Regge a fait l'hypothèse audacieuse de considérer une continuation analytique de l'amplitude de diffusion sur le plan complexe du moment cinétique, une variable qui, physiquement, ne prend que des valeurs discrètes. Durant la décennie suivante ces deux travaux ont  donné à la théorie de la matrice $S$ une impulsion incroyable. Elle a dominé la recherche dans le domaine des interactions fortes en atteignant presque le statut d'une religion. Le Temple était Berkeley et le Grand Prêtre G. Chew. On en reparlera à la section suivante mais ici je voudrais citer un travail important de Marcel Froissart fait pendant un séjour à Berkeley. En tenant compte de l'unitarité de la matrice $S$, Froissart a démontré que la section efficace totale de la collision entre deux particules scalaires, considérée comme fonction de l'énergie, ne peut pas croitre à l'infini plus vite que le carré du logarithme de l'énergie dans le système du centre de masse. C'est la “borne de Froissart”\cite{Froissart}. Dans la même période, Maurice Jacob, en visite à Brookhaven, a écrit un article avec G. C. Wick dans lequel ils introduisent le formalisme de l'hélicité pour décrire les amplitudes de diffusion des particules avec spin\cite{Jacob-W}.

\subsubsection { Le renouveau français} 
Nous avons vu aux sections précédentes que les contributions des chercheurs français en physique théorique des particules élémentaires durant les années 30 et 40 ont été d'un niveau plutôt médiocre. C'est seulement dans les années 50 qu'on observe les signes d'un certain renouveau.
\vskip 0.3cm
$\bullet$ {\bf Les trois mousquetaires.} Ce renouveau est surtout l'oeuvre de physiciens qui avaient effectué des séjours de formation à l'étranger. Les trois mousquetaires du titre sont Albert Messiah, Maurice Lévy et Louis Michel. Chacun d'eux, par son enseignement et sa recherche, a su attirer des jeunes et former des groupes de chercheurs actifs. Contrairement aux héros d'Alexandre Dumas, il n'y a pas eu de synergie réelle ou de collaboration scientifique entre eux. 

1) {\it Albert Messiah (1921-2013).} Il a interrompu ses études (X40) pour rejoindre la France Libre. Après la guerre il a étudié aux Etats-Unis (Institut de Princeton et Université de Rochester) et, à son retour en France, il a été recruté par Yves Rocard\footnote{Yves Rocard est un personnage atypique qui a joué un rôle très important dans ces développements même si il n'était pas un physicien des hautes énergies. Il a compris le grand intérêt de cette “nouvelle physique” et, de son double poste de directeur du laboratoire de physique de l'ENS et de conseiller scientifique du CEA, il a su attirer des jeunes scientifiques de talent. C'est encore lui qui a soutenu la construction du premier accélérateur linéaire à Orsay.} au CEA. Il a fait des travaux de recherche sur les propriétés statistiques de quarks mais il est surtout connu pour son enseignement de mécanique quantique, le premier cours structuré de cette discipline en France.

2) {\it Maurice Lévy (1922-2022).}  Il a passé sa thèse en France et il est parti d'abord en Angleterre et ensuite à l'Institut de Princeton. Dans son premier article il a cherché à déterminer les forces entre nucléons en tenant compte des effets dus à l'échange de deux pions\cite{Levy1}. De retour en France il a occupé des postes de Professeur à l'Université et il a été invité par Rocard pour créer un groupe de physique théorique à l'ENS. Ce groupe a essaimé à Orsay et à l'Université de Paris. Parmi les premiers membres du groupe on trouve A. Martin, Ph. Meyer et B. Jancovici. Le travail de recherche le plus connu de Lévy est celui du modèle
$\sigma$, écrit en 1960 en collaboration avec Gell-Mann\cite{Levy2}. Cet
article est remarquable à plusieurs titres: D'abord il est le premier modèle
explicite de la brisure spontanée de la symétrie chirale avec des mésons
$\pi$ comme bosons de Goldstone\footnote{Bien-sûr, le terme ``boson de
  Goldstone'' n'est pas utilisé, car l'article date de 1960. En fait, il
  précède de quelques jours l'article de Nambu.}. Ensuite, dans une note
ajoutée aux épreuves, il introduit la première notion d'universalité du
courant faible vectoriel dans la forme connue aujourd'hui, formulée dans le cadre de $SU(3)$ par N. Cabibbo en 1963. Enfin, les auteurs étudient aussi la limite dans laquelle la masse du champ $\sigma$ tend vers l'infini et obtiennent ainsi un modèle dans lequel les champs des mésons $\pi$ se transforment de façon non-linéaires sous les transformations chirales. Ce modèle $\sigma$ non-linéaire continue de jouer un rôle très important en physique théorique. Maurice Lévy est considéré, à juste titre, comme le principal artisan du
renouveau de la Physique des Hautes Énergies en France avec la création du “3ème
 Cycle de Physique Théorique 
Atomique et Nucléaire”, la première institution universitaire d'enseignement de la
physique moderne dans notre pays. Des générations de physiciens des hautes énergies, aussi bien des théoriciens que des expérimentateurs, ont été formés dans ce 3ème Cycle. C'est encore Lévy qui a créé l'\'Ecole de Physique à Cargèse\cite{Levy3}. 

3) {\it Louis Michel (1923-1999).} Après des études à l'\'Ecole Polytechnique (X43), Michel est allé à l'Université de Manchester où il a étudié sous la direction de L. Rosenfeld. Après sa thèse, obtenue en 1953, il a travaillé à l'Institut de Princeton et au groupe de physique théorique du CERN qui, à l'époque,  était installé à Copenhague. De retour en France il a enseigné à l'Université de Lille et, en 1960, il a créé un groupe de physique théorique à l'\'Ecole Polytechnique où il a attiré vers la physique des hautes énergies plusieurs élèves parmi les plus brillants de l'X\footnote{Une liste très incomplète des premières recrues contient  Cl. Bouchiat, H. Epstein et G. Flammand de la promotion 1953.}. Il a aussi contribué à la création de l'Institut des Hautes Etudes Scientifiques (I.H.E.S.) à Bures-sur-Yvette. L'oeuvre scientifique de Michel couvre un vaste domaine, de la phénoménologie des interactions faibles aux problèmes mathématiques liés à l'application de la théorie des groupes. Pour ce dernier sujet il a reçu la médaille Wigner en 1984. Pour sa thèse il a étudié le spectre des électrons émis lors de la désintégrations des leptons $\mu$ et il a montré que, pour une grande classe d'hamiltoniens, ce spectre dépend d'un seul paramètre, appelé “paramètre de Michel”\cite{Michel1}. En collaboration avec Cl. Bouchiat, il a étendu ces résultats en incluant les effets de polarisation et la violation de la parité\cite{Michel2}. Ils ont aussi étudié les contributions hadroniques dans le $g-2$ du muon\cite{Michel21}. Pendant son séjour à Copenhague Michel a introduit le concept de la “parité $G$”, le produit d'une transformation particule -- anti-particule $C$ avec une rotation de 180° dans l'espace d'isospin\cite{Michel3}. L'avantage est que tous les mésons $\pi$, chargés ou neutres, sont des états propres de $G$. En 1959, en collaboration avec V. Bargmann and V.L. Telegdi, ils ont obtenu l'équation qui décrit la précession de la polarisation d'une particule dans un champ électromagnétique\cite{Michel4}. 

Claude Bouchiat, élève de Michel, durant un séjour à l'Université de Princeton en 1958, a étudié les effets de recul dans la désintégration $\beta$ et la capture $K$ pour les transitions interdites\cite{Bouchiat1}.

\vskip 0.3cm
$\bullet$ {\bf L'enseignement de physique théorique en France. L'\'Ecole des Houches.} Jusqu'à la fin des années 50 l'enseignement de physique théorique était pratiquement absent du programme des universités et grandes écoles françaises. Il semble qu'il y avait à la place quelques “séminaires” assez confidentiels, formés de quelques jeunes autour d'un professeur. Dans la région parisienne on en connaissait au moins deux: celui de de Broglie et celui de Proca. Les deux s'occupaient des problèmes assez formels, sans contact réel avec des résultats expérimentaux: le premier surtout de questions épistémologiques et le deuxième des questions mathématiques de la théorie quantique des champs\footnote{Chaque séminaire avait ses fidèles, “la bande à de Broglie” et “la bande à Proca”. Parmi les physiciens que j'avais rencontrés, d'Espagnat m'avait dit d'avoir appartenu à la première et Ph. Meyer à la seconde.}. Malheureusement, Proca est mort en 1955 à l'âge de 58 ans.
\vskip 0.3cm
Dans ce désert académique une institution singulière a joué un rôle capital: il s'agit de l'\'Ecole de Physique Théorique des Houches fondée en 1950 par Cécile DeWitt-Morette (1922-2017). Elle avait débuté sa carrière au laboratoire de Joliot mais elle a dû partir à l'étranger afin d'étudier la physique mathématique. Pendant un séjour à l'Institut de Princeton elle avait épousé le physicien Américain Bryce DeWitt. Consciente des problèmes de l'enseignement en France elle a mis toute son énergie, qui était considérable, à la création d'une école d'été de physique théorique. Dans le projet qu'elle avait soumis au Ministère de l'\'Education Nationale elle expose les buts de cette \'Ecole avec une clarté remarquable.

{\it  PROJET de CREATION d’une ECOLE d’ETE

L’enseignement de la physique théorique est, en général, 
insuffisant. Dans ces conditions, il serait bon de mettre,
pendant l’été, à la disposition des étudiants et jeunes chercheurs 
français et étrangers, un enseignement de base leur permettant
d’aborder les problèmes de physique théorique moderne.
……..
Lorsqu’un enseignement régulier de la physique théorique sera 
organisé, cette école d’été subsisterait et dirigerait ses activités
vers des études plus spécialisées.}

Au départ,  chaque session aux Houches durait deux mois avec des cours intenses qui correspondaient à un bon semestre d'enseignement universitaire. La première session eut lieu en 1951 avec des cours, entre autres, de mécanique quantique par L. van Hove et de physique du solide par W. Kohn. En fait, le cours emblématique de mécanique quantique était enseigné pratiquement tous les ans durant la décennie 50-60 par des enseignants comme B. DeWitt, L. van Hove, J. Schwinger ou L. Schiff. L'\'Ecole compte parmi ses premiers élèves un grand nombre de physiciens de cette génération dont les futurs prix Nobel Alfred Kastler, Pierre-Gilles de Gennes, Georges Charpak, Claude Cohen-Tannoudji et Serge Haroche, pour ne mentionner que des Français. De même, parmi les enseignants on trouve tous les grands noms de notre discipline comme Fermi, Pauli, Feynman, Schwinger, Dyson, Glashow, Weinberg, Gross, Witten etc. Comme prévu dans le projet de Cécile DeWitt, au fur et à mesure que l'enseignement de base de physique théorique se répandait dans les universités françaises, l'enseignement aux Houches évoluait vers de sujets de recherche spécialisés.
\vskip 0.3cm
Pendant une bonne partie de la décennie l'\'Ecole des Houches était le seul endroit en France où un étudiant pouvait apprendre la physique moderne. Le cours de mécanique quantique de Messiah au CEA a débuté en 1955 et Lévy a créé le 3ème Cycle de physique théorique vers la fin des années 50. En 1962, l'année de mon inscription, le programme des cours était: mécanique quantique, par Bernard d'Espagnat, mécanique statistique, par Jacques Yvon, introduction à la
physique des particules élémentaires, par Philippe Meyer et physique
nucléaire, par Maurice Lévy. Il est difficile d'expliquer aux étudiants d'aujourd'hui que le
cours élémentaire de mécanique quantique, (le puits carré, l'oscillateur
harmonique, ou l'atome d'hydrogène), était enseigné pour la première fois au
niveau M2 et non pas en licence, mais en fait, avant la création de ce 3ème Cycle
 par Lévy, il n'était enseigné nulle part! Des générations de scientifiques
sortaient du système universitaire français dans l'ignorance totale, ou, pire,
avec des idées vagues et erronées, de ce chapitre fondamental de la physique. Le premier cours, on l'appelait Certificat à l'époque,  de mécanique quantique moderne à l'Université de Paris destiné aux étudiants de licence,  date de la fin des années 60.

\vskip 0.3cm
$\bullet$ {\bf Les premiers théoriciens français au CERN.} Le CERN est une des plus belles réussites scientifiques au monde, prouesse institutionnelle et prouesse technologique, un bel exemple de coopération internationale. Les scientifiques on su construire l'Europe de la Science bien avant que les politiques ne se penchent sur l'Europe de l'Economie\footnote{Un des premiers scientifiques à proposer la création. d'un centre européen de recherche en physique des hautes énergies fut L. de Broglie lors de la conférence européenne de la culture à Lausanne en décembre 1949.}\cite{CERN1}. Le CERN est un
 centre de physique expérimentale mais, dès les premières années de son existence, il s'est doté d'un groupe actif de physique théorique\footnote{En fait, sur ce point le CERN n'a pas suivi l'exemple Américain. Brookhaven, l'analogue mais aussi le rival du CERN pendant les premières années, n'a jamais acquis un grand groupe de physique théorique. Qui plus est, au CERN le rôle des théoriciens était clairement défini: même s'ils étaient encouragés à collaborer avec les expérimentateurs, ceci n'a jamais été une obligation. L'excellence dans la recherche était le critère essentiel. Parmi les premières recrues ainsi que parmi les membres permanents plus tard, on trouve des spécialistes de physique mathématique.}. Entre 1952 et 1957, les années de construction à Genève, les théoriciens étaient affectés au Centre de Physique Théorique “Niels Bohr” à Copenhague et Bohr a été le premier Directeur du groupe. Deux théoriciens Français ont été boursiers du CERN à Copenhague: Louis Michel et Bernard d'Espagnat.  
 
Le premier Directeur Général du CERN, en 1954,  a été Felix Bloch. Il était sensé aller à Genève afin de superviser la construction du centre et il a obtenu du Conseil la permission d'inviter un petit groupe de physiciens. Il a d'abord offert le poste à Anatole Abragam qui, comme Bloch, était un spécialiste du magnétisme nucléaire. A l'époque Abragam dirigeait un groupe à Saclay et n'a pas voulu déménager de façon permanente à Genève. Le premier théoricien recruté en tant que membre au CERN fut d'Espagnat (Oct. 1954) suivi en Janvier 1955 par Jacques Prentki. Les deux venaient du laboratoire de Leprince-Ringuet de l'\'Ecole Polytechnique et le travail qu'ils ont fait ensemble est, probablement, le plus significatif de tout ce qui a été fait à la Division Théorie à Genève pendant cette décennie. A partir de 1953, plusieurs théoriciens avaient essayé d'élargir la symétrie d'isospin afin d'inclure la conservation de l'étrangeté. Dans cet effort nous voulons citer une série d'articles de d'Espagnat et Prentki\cite{Esp-Pr1}. Ils ont prédit  l'existence d'un méson $0^-$ isoscalaire, bien avant la découverte de $\eta$ et $\eta'$ et ils ont abordé pour la première fois la question générale de symétrie pour toutes les interactions.  En 1962 ils adaptèrent ce travail au groupe de symétrie  $SU(3)$\cite{Esp-Pr2} et écrivirent le courant faible sous une forme qui ressemble à celle introduite par N. Cabibbo en 1963 avec un angle qu'ils appelèrent $\alpha$. Mais ils n'ont pas exploité ce principe à cause, paraît-il, d'un résultat expérimental erroné qui indiquait une violation de la règle $\Delta Q = \Delta S$ dans les désintégrations des mésons $K$. Je voudrais aussi mentionner un travail de R. Omnès qui était boursier au CERN. En 1958 il a établi une équation intégrale, connue sous le nom d'équation de Muskhelishvili-Omnes, qui relie des quantités mesurables d'une réaction qui contient une paire de mésons, par exemple deux pions, dans l'état final, à l'amplitude de diffusion élastique à basse énergie de ces mésons\cite{Omnes}.

\subsection{1960 - 1970 : L'hégémonie de la matrice $S$ - Vers la construction du Modèle Standard}
Au cours de cette décennie la recherche en physique théorique des hautes énergies fut marquée par deux courants principaux: l'étude des interactions fortes dans le cadre de la théorie de la matrice $S$ d'un côté et l'extension des propriétés de symétrie à des groupes au delà du groupe d'isospin de l'autre. En parallèle  on avait quelques travaux, presque confidentiels, par des chercheurs isolés. Certains commençaient déjà à jalonner le chemin vers la construction du Modèle Standard. 

C'est aussi la décennie durant laquelle la recherche s'est bien développée en France. Ce développement s'est fait en grande partie dans le cadre du CNRS grâce au système des laboratoires associés.  Pour la physique des hautes énergies nous constatons une particularité notable:  les centres de physique théorique sont distincts de ceux de la physique expérimentale, contrairement à ce qui se passe dans d'autres disciplines, telles la physique atomique ou la physique de la matière condensée. Peut-être à cause de cette séparation, les laboratoires de physique théorique sont davantage “pluridisciplinaires” avec des spécialistes de la mécanique statistique et de la physique des transitions de phase qui côtoient ceux des particules élémentaires. Naturellement -- nous sommes en France -- nous observons une grande concentration dans la région Parisienne (Orsay, Saclay, \'Ecole Polytechnique, Jussieu) avec quelque exceptions, surtout à Marseille et un peu à Bordeaux. L'émergence des centres en province débute vers la fin de la décennie. 

\subsubsection{La théorie de la matrice $S$ et les interactions fortes.} C'était de loin le courant dominant.   Cette approche qui, d'un optimisme exagéré, est tombée à un oubli tout aussi injustifié\footnote{Aujourd'hui certains résultats et techniques de la théorie de la matrice $S$, comme par exemple la méthode du “bootstrap”, sont ressuscités.},  a produit plusieurs résultats intéressants, parfois par des chercheurs français. 

La théorie de la matrice $S$ postule des propriétés d'analyticité pour les amplitudes de diffusion. 
J. Bros et H. Epstein, en collaboration surtout avec V. Glaser, ont démontré que, dans le cadre de la théorie quantique des champs, l'amplitude de diffusion, considérée comme fonction du carré de l'énergie dans le centre de masse $s$ et de celui du moment de transfert $t$, admet un prolongement analytique en $s$ et $t$ complexes dans un domaine qui contient la région physique. En plus, cette fonction analytique satisfait à la propriété de croisement qui consiste à échanger entre des particules entrantes et sortantes\cite{Br-Ep-Gl}. Pour écrire des relations de dispersion la propriété d'analyticité ne suffit pas, il faut encore que l'amplitude soit bornée à l'infini par un polynôme dans toutes les directions de $s$ et $t$ complexes. Afin d'obtenir des  bornes de ce genre on utilise 
la relation de l'unitarité, qui est une relation non-linéaire dérivée du principe de conservation des probabilités. La borne de Froissart\cite{Froissart} en est un exemple.
 Les résultats les plus complets obtenus grâce aux contraintes imposées par l'unitarité sont dus à A. Martin. Pendant les années 60 et 70 il a créé au CERN la meilleure école sur ce problème, a guidé le travail de plusieurs jeunes théoriciens et est devenu l'expert reconnu sur le plan mondial. Il a démontré des bornes rigoureuses à la Froissart pour les sections efficaces hadroniques, il a élargi le domaine d'analyticité et il a obtenu des résultats très précis pour l'amplitude de diffusion entre deux mésons $\pi$\cite{Martin}. 
 
 En France cette ligne de recherche a été poursuivie surtout à Orsay et à Saclay. B. Diu, J.L. Gervais et H. Rubinstein\cite{Diu-Rub} au laboratoire d'Orsay ont appliqué le programme du bootstrap\footnote{En général, nous supposons que si nous connaissons toutes les forces entre deux particules, nous pouvons calculer toutes les propriétés de l'amplitude de diffusion. L'hypothèse initiale du bootstrap est une approximation, un peu simpliste, de cette idée. On suppose que les forces produites par l'échange d'une particule dans la voie croisée sont suffisantes pour produire l'état de cette particule dans la voie directe. D'où le terme “bootstrap” du conte de M\"unchhausen. Une particule produit les forces qui la créent. On obtient ainsi un système d'équations d'auto-cohérence. L'approximation d'une particule est trop simple, mais l'idée générale  est devenue aujourd'hui un outil puissant aussi bien en théorie des champs qu'en mécanique statistique.}, initialement proposé pour le méson $\rho$, à d'autres particules, comme le $K^*$ et ils ont montré que l'accord avec l'expérience est, au mieux, qualitatif. A Saclay, l'équipe de G. Cohen-Tannoudji, A. Morel et H. Navelet a présenté une étude cinématique complète des amplitudes de diffusion à deux corps dans la base d'hélicité\cite{COMONAV} qui a été très utile pour l'analyse de résultats expérimentaux.
 
Dans le sillage de la théorie de la matrice $S$, un des articles les plus influents de cette période est celui de G. Veneziano. Il a postulé une expression explicite pour l'amplitude de diffusion sans se poser la question de savoir si elle découle d'un modèle dynamique. Cette expression très simple sous forme d'un rapport de fonctions $\Gamma$, a des propriétés presque magiques: la symétrie de croisement, un comportement en accord avec la théorie de Regge et, le plus important, une propriété de {\it dualité} qui, en gros, relie l'amplitude ainsi que le spectre des particules qu'on observe dans une expérience de diffusion aux forces produites par l'échange des particules à la Yukawa. La connexion de ce modèle avec la théorie d'une corde vibrante est apparue vers la fin de la décennie et elle a ouvert des nouvelles voies de recherche, telles les modèles duaux et la théorie des cordes, qui sont encore en pleine expansion.

\subsubsection{Les symétries unitaires et l'arrivée des quarks.}\label{SU3}  La découverte des particules étranges indiquait une symétrie des interactions fortes plus large que le groupe $SU(2)$ d'isospin et cette recherche était déjà initiée la décennie précédente.  La solution correcte, basée sur le groupe $SU(3)$, fut trouvée en 1961 par M. Gell-Mann et Y. Ne'eman. Malgré les apparences, le passage de $SU(2)$ à $SU(3)$ était tout sauf évident. En fait, sur le spectre des hadrons, ces deux groupes se manifestent de façon différente. $SU(2)$ est facile à deviner parce que nous observons un doublet de nucléons, le proton et le neutron. L'élargissement simple consisterait à ajouter un troisième membre et de passer d'un doublet à un triplet. Un modèle de ce genre fut proposé par S. Sakata en 1959 mais ce n'est pas la solution choisie par la Nature. En réalité le doublet de $SU(2)$ devient un octet de $SU(3)$. Cette absence apparente de triplet a conduit Gell-Mann et G. Zweig en 1964 a postuler l'existence d'une nouvelle strate dans la structure de la matière. Les hadrons, comme le proton, le neutron, les mésons etc, ne seraient plus de particules élémentaires mais composés à partir de constituants, appelés “quarks”\footnote{Gell-Mann a trouvé ce mot dans “Finnengans Wake” de James Joyce.} par Gell-Mann. En 1964 le nombre de quarks nécessaires pour reproduire toutes les particules connues  était trois (qu'on désigne par $u$, $d$ et $s$) qui forment le triplet manquant.  Aujourd'hui nous en connaissons au total six, les trois de Gell-Mann et Zweig et trois nouveaux, le $c$, le $b$ et le $t$. Ils ont tous un spin égal à 1/2 et chacun porte un nombre quantique distinct qui, dans le jargon des physiciens, est appelé {\it saveur}. Les baryons (proton, neutron, etc.) sont des états liés de trois quarks tandis que les mésons, tels le $\pi$ ou le $K$, des états liés d'une paire quark -- anti-quark. Les masses des quarks couvrent un vaste domaine, $u$ et $d$ sont légers et $t$ est la plus lourde des particules élémentaires. Les charges électriques sont des fractions de la charge du proton\footnote{Cette propriété des quarks d'avoir des charges fractionnaires apparut étrange et des modèles avec charges entières furent aussi proposés, en particulier par des chercheurs au CERN\cite{Bacry-N-VH}. Aujourd'hui les résultats expérimentaux sont en faveur du modèle initial avec des charges fractionnaires.}. En fait, le modèle des quarks s'est avéré être un peu plus compliqué. Afin de rendre compte du principe de Pauli qui interdit à deux quarks identiques d'occuper le même état quantique, nous supposons que chaque saveur de quark existe en trois espèces différentes qu'on appelle  {\it couleurs}\footnote{Ces termes “saveur” et “couleur” n'ont aucune relation avec les sens ordinaires de ces mots.}. Mais cela n'est pas identique à un modèle à 18 quarks parce que chaque état physique contient les trois couleurs en proportions égales.

L'hypothèse des quarks a déclenché une intense activité de recherche, aussi bien expérimentale -- un des buts de la construction des anneaux de collision ISR au CERN était de mettre en évidence les quarks -- que théorique, à laquelle des chercheurs français ont participé\cite{Orsay-quarks}.

Un résultat indirect du travail de Gell-Mann a été de faire connaître la théorie des groupes aux physiciens des hautes énergies. Certains théoriciens français, qui avaient souvent une bonne culture mathématique, ont contribué à ce processus et ont obtenu des résultats importants\cite{Itz-etal}.

\subsubsection{Les interactions faibles}  
$\bullet$ {\bf Deux grandes réussites expérimentales.} 

 En 1962 une équipe de l'Université de Columbia, travaillant auprès de l'accélérateur de Brookhaven, a établi l'existence de deux espèces distinctes de neutrinos, un associé à l'électron et un autre au muon\footnote{Ce résultat était anticipé par des théoriciens, par ex. Schwinger. Aujourd'hui nous connaissons l'existence d'un troisième associé au lepton $\tau$. Nous avons aussi observé des phénomènes d'oscillation entre ces trois espèces.}. C'était la première expérience à obtenir des résultats avec un faisceau de neutrinos et la première à utiliser des chambres à étincelles comme détecteur\footnote{C'était aussi la première grande compétition entre CERN et Brookhaven. L'accélérateur au CERN avait une année d'avance sur celui de Brookhaven, mais les expérimentateurs en Europe n'ont pas su en profiter  et ils se sont contentés de confirmer le résultat de Brookhaven. En revanche, l'expérience du CERN était supérieure sur le plan technologique et a établi les standards pour toutes les futures expériences neutrino.}. 

 En 1964 une équipe de l'Université de Princeton a mis en évidence des effets de violation de la symétrie $CP$ -- le produit de l'opérateur d'échange particule-antiparticule avec l'opérateur parité -- dans les désintégrations des mésons $K$ neutres. Nous avons déjà vu que toute théorie quantique des champs qui satisfait aux axiomes de Wightman est invariante sous le produit $CPT$. Par conséquent, le résultat de cette expérience peut être interprété comme une violation de la symétrie par renversement du temps. 

Ces deux expériences ont profondément influencé la recherche théorique jusqu'à nos jours. 
\vskip 0.3cm
$\bullet$ {\bf La théorie de Cabibbo.} La théorie $V-A$ que nous avons vue à la section \ref{50-60-Int} postule que les courants des interactions faibles font partie de triplets de la symétrie $SU(2)$ d'isospin. d'Espagnat et Prentki ont essayé de généraliser cette propriété à des symétries plus larges et, en particulier, à $SU(3)$. La solution fut trouvée par N. Cabibbo en 1963 qui montra que la généralisation correcte de $SU(2)$ à $SU(3)$ implique que les courants qui apparaissent dans les interactions faibles appartiennent à un octet. Dans cette formulation la propriété de l'universalité des interactions faibles sous forme algébrique dévient manifeste avec l'introduction de l'angle connu sous le nom d'“angle de Cabibbo”. C'est un des articles fondateurs de la théorie des interactions faibles.

En 1964, M. Ademollo et R. Gatto et, indépendament, Cl. Bouchiat et Ph. Meyer\cite{Bouch-Mey1} ont montré que, sous les hypothèses de la théorie de Cabibbo, les constantes de couplage du courant vectoriel des interactions faibles ne sont pas renormalisées au premier ordre de l'interaction qui brise $SU(3)$.
\vskip 0.3cm
$\bullet$ {\bf L'algèbre des courants et la symétrie chirale.} C'est probablement la plus importante contribution de Gell-Mann, celle à laquelle il a consacré ses efforts pendant des longues années. L'objet était de découvrir des symétries, qui pourraient être cachées, des interactions fortes, en étudiant les propriétés des courants des interactions électromagnétiques et faibles. Nous pouvons suivre le cheminement de sa pensée à travers une série d'articles, de la théorie $V-A$ de 1957, au modèle $\sigma$ de 1960, jusqu'à l'algèbre d'opérateurs sur le cône de lumière des années 70. Il a publié une trentaine d'articles, avec plusieurs collaborateurs\footnote{Parmi ces articles deux sont publiés durant son séjour à Orsay en 1960. Le premier est le modèle $\sigma$ avec M. Lévy\cite{Levy2} et le deuxième, sur les propriétés de renormalisation du courant axial, avec J. Bernstein et L. Michel\cite{Gell-Mich}.}. Nous avons vu dans la section \ref{50-60-Int} la théorie du courant vectoriel conservé (CVC) qui postule que la partie vectorielle du courant hadronique des interactions faibles fait partie du triplet des courants conservés de la symétrie d'isospin\footnote{Si nous considérons la symétrie élargie que nous avons introduite dans la section \ref{SU3}, le courant vectoriel fait partie de l'octet des courants conservés de $SU(3)$.}. On avait signalé dans la section  \ref{50-60-Int} que l'hypothèse du CVC résout seulement la moitié du problème parce qu'elle laisse de côté la partie axiale. C'était la partie la plus difficile et c'est le grand mérite de Gell-Mann d'avoir trouvé la solution. 
Le résultat le plus important de son travail est la découverte de la symétrie chirale $SU(3)\times SU(3)$ qui est générée par les courants vectoriels et axiaux. La confirmation, pour la partie $SU(2)\times SU(2)$ de cette symétrie, fut obtenue en 1965 par la relation de S. L. Adler et W. I. Weisberger qui relie les constantes de l'interaction forte, telle la constante de couplage pion-nucléon, avec le facteur de forme du courant axial des interactions faibles. Le point important est que cette symétrie n'est pas manifeste sur le spectre des particules. Les hadrons se classent en représentations d'isospin ou de $SU(3)$, mais pas en celles de 
$SU(3)\times SU(3)$. Les nucléons n'ont pas de partenaires de parité opposée. Comprendre le mécanisme de la brisure de la symétrie chirale a été un grand défi, mais aussi une source intarissable de découvertes. 

\subsubsection{Recherches hors des sentiers battus.} Par définition, les recherches dans ce domaine sont inclassables. La liste qui suit inclut plusieurs aspects de la théorie quantique des champs. C'est une sélection personnelle et arbitraire qui parfois dépasse cette décennie. 
\vskip 0.3cm
$\bullet$ {\bf L'électrodynamique quantique.} Un sujet très minoritaire, avec des contributions des chercheurs français, pas forcement liées les unes aux autres, mais souvent importantes. 

En 1960 M. Froissart et R. Stora ont étudié les effets de dépolarisation d'un faisceau de protons polarisés injecté dans un synchro-cyclotron dus aux inhomogénéités du champ magnétique\cite{Fr-St}. 

En 1969 C. De Callan, R. Stora et W. Zimmermann ont montré que l'électrodynamique peut être considérée comme limite d'une théorie avec un photon massif lorsque la masse tend vers zéro\cite{Stora1}.

En 1970 E. Brézin, Cl. Itzykson et J. Zinn-Justin ont montré que la somme d'une classe de diagrammes en échelle de l'électrodynamique peut donner une généralisation relativiste de la formule de Balmer qui inclut les effets de recul. Par ailleurs Brézin et Itzykson ont estimé la production de paires $e^+-e^-$ dans le vide par des effets non-linéaires de l'électrodynamique\cite{Itzyk2}.

Entre 1968 et 1971 E. De Rafael et ses collaborateurs ont publié une série d'articles sur le calcul du Lambshift\cite{deRaf1} ainsi que du $g-2$ de l'électron et du muon\cite{deRaf}. Ces calculs ont eu une grande influence jusqu'à nos jours.
\vskip 0.3cm
$\bullet$ {\bf Aspects mathématiques de la théorie quantique des champs.} C'est un sujet qui a toujours attiré des chercheurs français. 

En 1964 R. Haag et D. Kastler ont développé une formulation algébrique de la théorie quantique des champs\cite{Kastler1}. En particulier ils ont démontré qu'une théorie des champs peut être définie à partir de l'algèbre satisfaite par les opérateurs qui représentent les observables. 

En 1969 H. Abarbanel et Cl. Itzykson ont présenté une méthode pour obtenir le développement eikonal dans le cadre de la théorie quantique des champs relativiste\cite{Itzyk1}.

D'habitude on présente le programme de renormalisation dans l'espace des impulsions qui est celui qu'on obtient par le calcul des diagrammes de Feynman. En 1973 Epstein et Glaser ont reformulé ce programme directement dans l'espace des configurations $x$\cite{Epst-Gl} qui rend le contenu physique plus transparent. La même année Epstein, Glaser et Stora ont étudié les propriétés mathématiques de la fonction de Green à $N$ points en théorie quantique des champs\cite{Stora2}.

\vskip 0.3cm
$\bullet$ {\bf Les anomalies des identités de Ward en théorie quantique des champs.} En théorie classique des champs l'invariance sous un groupe de transformations continues implique l'existence d'un courant conservé. C'est la conséquence du théorème de Noether. Cependant, au niveau quantique, la conservation des tous les courants d'une symétrie n'est pas toujours garantie. Un exemple bien connu est l'anomalie ABJ (Adler-Bell-Jackiw) du courant axial qui montre que, dans un modèle invariant sous des transformations chirales, nous ne pouvons pas toujours imposer la conservation simultanée des courants vectoriels et axiaux. Une généralisation de ces équations de conservation a été obtenue par un groupe à Orsay\cite{Amati1}. Cet effet, qui semble purement technique, s'est avéré être d'une grande importance, pas seulement pour la construction du Modèle Standard, mais aussi pour des tentatives d'aller au-delà. 

\vskip 0.3cm
$\bullet$ {\bf Brisure spontanée d'une symétrie.} Nous avons l'habitude de chercher aux problèmes symétriques des solutions
symétriques. Pour citer Pierre Curie:\cite{SSB1} 
{\it “Lorsque certaines causes produisent
certains effets, les éléments de symétrie des causes doivent se retrouver dans
les effets produits.”} Or, nous savons déjà en physique classique des exemples des systèmes pour lesquels l'état fondamental peut avoir une symétrie plus petite que celle des équations dynamiques du système. C'est le phénomène de “brisure spontanée” d'une symétrie\footnote{Pour un exposé simple de ce phénomène voir le Chapitre 5 du livre de la SFP\cite{LivreSFP}.}. Il est omniprésent en physique classique, en physique de la matière condensée ainsi que dans les phénomènes de transition de phase, mais son application en physique des hautes énergies date de 1960. Nous avons vu que la symétrie chirale n'est  pas  manifeste dans le spectre des hadrons. Y. Nambu a remarqué qu'elle pourrait être spontanément brisée et, en 1961, J. Goldstone a formulé un théorème, qui porte son nom, selon lequel à la brisure spontanée d'une symétrie continue correspond une particule de masse nulle, le “boson de Goldstone”. Une démonstration mathématique de ce théorème est due à l'équipe de Marseille\cite{Kastler2}. Les bosons de Goldstone de la symétrie chirale sont les mesons $\pi$ (ou les mésons $K$ si nous considérons $SU(3)\times SU(3)$). Comme la symétrie n'est pas exacte, leurs masses ne sont pas exactement égales à zéro, mais elles sont relativement petites dans l'échelle des masses hadroniques. On les appelle des bosons “pseudo-Goldstone”.

\vskip 0.3cm
$\bullet$ {\bf Les fondements de la mécanique quantique.} Avant de quitter cette section je voudrais mentionner un travail qui fut certainement hors de tout sentier battu; celui de J. Bell et ses célèbres inégalités. A strictement parler il ne s'agit pas d'un travail de physique des hautes énergies, mais il a contribué de façon essentielle à notre meilleure compréhension de la mécanique quantique et, surtout, il a ouvert un nouveau domaine de recherche. Nous avons vu à la section \ref{MQ} la suggestion  d'une onde pilote faite par de Broglie. Au début des années 50 cette idée fut reprise par le physicien britannique D. Bohm qui proposa un modèle “à variables cachées”. Il décrit le mouvement des électrons suivant des trajectoires classiques dont les coordonnées ne sont pas observables par des mesures macroscopiques; ce sont des variables cachées. Afin de reproduire les résultats de la mécanique quantique, Bohm introduit des corrélations non-locales entre ces variables\footnote{On lit souvent que cette non-localité de la théorie de Bohm est équivalente à la non-localité de la fonction d'onde de la mécanique quantique, mais ceci n'est pas correct. Sans entrer dans des détails, je rappelle que, pour la seconde, nous pouvons facilement écrire une généralisation compatible avec la relativité restreinte, mais pas pour la première. Il se peut que l'extension relativiste d'une théorie non-locale à variables cachées doive être cherchée dans le cadre d'une théorie quantique de la gravitation.}. Entre 1964 et 1966 J. Bell obtint deux résultats importants: le premier est technique: il montra qu'un théorème, attribué à von Neumann, sur l'impossibilité mathématique d'une théorie à variables cachées reproduisant les prédictions de la mécanique quantique, était en fait incorrect. La démonstration était basée sur des hypothèses illégitimes concernant les états d'un système physique. Le deuxième, qui est le plus important, montre que la question est, en fait, physique, c'est à dire elle peut être résolue par l'expérience. Avec ses fameuses inégalités il montre que toute théorie locale à variables cachées a forcement des prédictions différentes de celles de la mécanique quantique. Ce résultat a profondément affecté notre perception sur la nature fondamentale de la mécanique quantique. En plus, il a ouvert un nouveau champ de recherches expérimentales,  initiées en France par A. Aspect et ses collaborateurs\cite{Aspect}, qui fut couronné par le prix Nobel 2022.

\vskip 0.3cm
\subsubsection { Vers le Modèle Standard - La théorie électrofaible.} La recherche qui a abouti à la construction du Modèle Standard a suivi plusieurs voies parallèles, sans relation apparente entre elles. Néanmoins, le succès final est dû à la réussite simultanée de toutes. C'était l'aboutissement des efforts de certains chercheurs isolés. Les réussites individuelles passaient le plus souvent inaperçues. Les protagonistes n'étaient pas toujours au courant les uns des travaux des autres. Plusieurs résultats ont dû être redécouverts plus d'une fois\footnote{Pour un exposé historique voir \cite{Rise-SM}}. 
\vskip 0.3cm
$\bullet$ {\bf La théorie de Yang-Mills.} La motivation de Yang et Mills était de trouver une théorie pour les interactions fortes et le groupe était $SU(2)$ pour l'isospin. Cependant la possibilité d'une application aux interactions électromagnétiques et faibles avait été reconnue très vite, en particulier par S. Bludman et J. Schwinger. En 1960 Gell-Mann et Glashow ont généralisé l'invariance de jauge à des groupes quelconques et ils ont étudié les conséquences pour les interactions faibles. Le travail le plus important de cette période date de 1961. C'est un modèle de Glashow basé sur le groupe $SU(2)\times U(1)$ pour les interactions électromagnétiques et faibles, dans lequel le photon est une combinaison linéaire du boson de jauge de $U(1)$ et de celui de la composante neutre de $SU(2)$. 
\vskip 0.3cm
$\bullet$ {\bf Les problèmes de masse.} L'invariance de jauge semble imposer aux bosons vectoriels d'avoir des masses nulles. C'est le premier -- et le plus important -- “problème de masse”. Il a retardé l'application des théories de Yang-Mills pendant de longues années et c'est seulement en 1964 que la solution fut trouvée. C'est le mécanisme bien connu de Brout-Englert-Higgs (BEH)\footnote{Il existe plusieurs exposés simples de ce mécanisme. Voir, par exemple, le Chapitre 5 de la référence\cite{LivreSFP}, ou encore \cite{Ilio1}.}. Tout comme les théories de Yang-Mills, ce mécanisme était proposé en vue d'une application aux interactions fortes. La théorie envisagée était la symétrie $SU(3)$ de Gell-Mann et Ne'eman, écrite comme une théorie de Yang-Mills spontanément brisée afin de donner une masse aux bosons de jauge. Ces derniers étaient sensés être les bosons vectoriels $\rho$, $K^*$ etc. L'application aux interactions faibles a été découverte plus tard.

En fait, le problème de masse pour une théorie de jauge des interactions électromagnétiques et faibles a aussi un deuxième volet: nous savons que les courants qui interviennent dans la théorie de Fermi sont de la forme $V-A$, la différence entre un courant vectoriel et un courant axiale. Dans une théorie de Yang-Mills ces courants sont ceux de la symétrie de jauge. Si nous essayons d'exprimer ces courants en termes de champs des fermions qui participent aux interactions faibles, l'électron, le muon, les neutrinos ou les quarks, nous trouvons que la forme $V-A$ implique que tous ces fermions doivent aussi avoir des masses nulles\cite{Ilio1}. C'est le deuxième “problème de masse”. 
\vskip 0.3cm
$\bullet$ {\bf La synthèse - Le modèle de Weinberg-Salam.} S. Weinberg, en 1967, et A. Salam en 1968 ont trouvé une solution à ces problèmes de masse.  Ils ont combiné le modèle $SU(2)\times U(1)$ de Glashow avec une version du mécanisme BEH et ils ont montré que ce dernier pourrait être à l'origine d'une série de phénomènes remarquables. (i) Produire le mélange postulé par Glashow entre les bosons de jauge de $U(1)$ et la composante neutre de $SU(2)$ pour donner naissance au photon et un boson neutre. (ii) Donner des masses à trois bosons vectoriels laissant le photon de masse nulle et expliquer ainsi la séparation entre les interactions électromagnétiques et faibles observée à basse énergie. (iii) Donner des masses non-nulles aux fermions.

C'était une solution partielle parce que, pour des raisons physiques que nous allons expliquer plus loin, elle ne s'appliquait qu'aux leptons -- l'électron, le muon et les neutrinos -- et pas aux quarks.
\vskip 0.3cm
$\bullet$ {\bf Les divergences de la théorie de Fermi.} En parallèle et de façon totalement indépendante, une autre histoire s'est déroulée durant la même époque. La théorie de Fermi n'est pas renormalisable, on ne sait calculer que les termes du premier ordre de la théorie des perturbations. La constante de couplage $G_F$ a des dimensions [M]$^{-2}$ et sa valeur, exprimée en unités de la masse du proton,  est $G_F\approx 10^{-5} m_p^{-2}$. Afin d'estimer l'ordre de grandeur des termes suivants dans le développement perturbatif, c'est à dire des termes en $G_F^2$, $G_F^3$, etc, on est obligé d'introduire un paramètre $\Lambda$ avec des dimensions d'une masse qu'on appelle “cut-off”. Ainsi, les termes du premier ordre sont proportionnels à $G_F$, ceux du deuxième à $G_F^2 \Lambda^2$, les suivants à $G_F^3\Lambda^4$ etc. Nous voyons donc que nous ne pouvons avoir confiance aux termes du premier ordre que si $G_F\Lambda^2 <1$, ce qui donne $\Lambda < $ 300 GeV. Pour les physiciens des années 60, une énergie de 300 GeV était pratiquement infinie, c'est la raison pour laquelle très peu de gens se préoccupaient de ces divergences. Cependant, en 1967, B.L. Joffe et E.P. Shabalin en URSS ont remarqué que la limite de 300 GeV peut être abaissé de façon drastique. Les interactions faibles violent la conservation de la parité et de l'étrangeté. L'absence d'effets de ce genre dans les expériences avec interactions fortes, par exemple en physique nucléaire, donne une limite beaucoup plus contraignante, à savoir $\Lambda < $ 3 GeV. 

On s'attendrait à une forte réaction de la communauté à ce résultat parce qu'il implique que des effets des interactions faibles devraient être visibles aux interactions hadroniques à l'échelle de 3 GeV -- l'énergie des accélérateurs atteignait alors 30 CeV -- contrairement à toute évidence. Or, très peu de gens l'ont remarqué\footnote{Pour la grande majorité des physiciens, le problème de la physique des hautes énergies était celui des interactions fortes. Toute autre étude était considérée comme une perte de temps. Il peut paraître paradoxal qu'une révolution dans notre compréhension des interactions fondamentales soit venue par l'étude de la plus faible d'entre elles, mais ce n'est pas la première fois. Nous avons d'autres exemples qui montrent l'importance potentielle des “petits” effets.} et la réponse, qui fut trouvée en deux étapes, est passée pratiquement inaperçue. 

La première étape concernait les effets de violation de la parité et/ou de l'étrangeté dans les interactions fortes qui étaient dus aux divergences qu'on appelait “dominantes”. La solution fut trouvée par une équipe du CERN\cite{BIP} qui a montré que ces effets sont absents si les termes qui violent l'invariance chirale des interactions fortes ont la forme d'un terme de masse des quarks. 

\vskip 0.3cm
$\bullet$ {\bf Les courants neutres et le charme.} Le succès de la première étape n'a pas apporté des changements importants dans notre perception de la théorie des interactions fortes. La brisure de l'invariance chirale par les termes de masse des quarks était généralement admise. Ce ne fut pas le cas de la deuxième étape. Elle concernait des divergences appelées “sous-dominantes” qui auraient comme effet l'apparition des désintégrations du genre $K^0_L \rightarrow \mu^++\mu^-$, c'est à dire des effets de courants neutres avec violation de l'étrangeté, pour lesquelles les limites expérimentales étaient déjà assez strictes. A l'époque, l'absence de ces désintégrations était interprétée comme absence totale des courants neutres et le modèle des interactions faibles prévoyait uniquement l'existence des bosons vectoriels chargés. La solution de ce problème eut des conséquences plus radicales. Le point essentiel était de réaliser que, expérimentalement, seuls les processus de courants neutres avec $\Delta S\neq 0$, c'est à dire avec changement d'étrangeté, étaient interdits; ceux avec $\Delta S= 0$ ne l'étaient point. Par conséquent, il ne fallait pas chercher à éliminer les courants neutres à tous les processus, mais seulement à ceux qui changent l'étrangeté. Or, si on regardait les interactions faibles entre leptons, les divergences sous-dominantes ne généraient pas des processus avec changement des nombres leptoniques. Cette différence de comportement était due à une dissymétrie entre les leptons et les hadrons: on avait quatre leptons mais seulement trois quarks. Maintenant la solution apparaît évidente: il fallait postuler l'existence d'un quatrième quark\footnote{Appellé “charm” $c$. En 1973  M. Kobayashi et T. Maskawa ont étendu le nombre des quarks de 4 à 6 afin d'avoir un mécanisme naturel pour la violation de la symétrie $CP$.} et rétablir ainsi la symétrie leptons -- hadrons\cite{GIM}. Ce travail devait permettre l'application du modèle de Weinberg et Salam aux quarks en levant le dernier obstacle phénoménologique pour la formulation des interactions électrofaibles comme une théorie de Yang-Mills, même si, dans la pratique, l'histoire fut un peu plus compliquée\cite{Rise-SM}.

L'existence d'un quatrième quark implique une extension de la symétrie approximative des interactions fortes de $SU(3)$ à $SU(4)$\footnote{Ce genre d'extension avait été proposé auparavant comme une généralisation de $SU(3)$ sans justification dynamique particulière, voir p. ex.\cite{Amati2}.} avec une multitude de nouveaux hadrons à la composition desquels ce quark participe. 

\vskip 0.3cm
$\bullet$ {\bf La théorie quantique des champs de Yang-Mills.} La troisième voie vers la construction du Modèle Standard a été l'étude des propriétés de renormalisation de la théorie de Yang-Mills. Elle a été initiée par M. Veltman au milieu des années 60, joint un peu plus tard par son étudiant G. 't Hooft. Ce fut un travail long, ardu et compliqué, pas seulement à cause de la nature du travail, mais aussi parce qu'ils ont été obligés d'inventer et parfois de re-inventer, plusieurs techniques nouvelles\cite{Rise-SM}. Le problème du choix de jauge n'était pas évident, les règles de Feynman n'étaient pas connues et ils n'étaient pas au courant de résultats de Brout, Englert et Higgs. Leur réussite fut un vrai tour de force. 

\vskip 0.3cm
\subsubsection { Vers le Modèle Standard - Les interactions fortes.}\label{SLAC}
 Les interactions fortes sont entrées en scène dans les années 30 avec la physique nucléaire, mais leur développement a suivi celui des accélérateurs. Pendant des années l'outil principal était l'étude des collisions hadroniques et, grâce à elles, nous avons découvert un très grand nombre d'états qu'on décrit comme “résonances”\footnote{Rien que pour le système pion-nucléon, la Table des Particules contient une trentaine de résonances avec masses entre 1 et 2.5 GeV et spins jusqu'à 11/2.}. Nous n'avions aucun moyen d'expliquer cette prolifération à partir d'équations fondamentales. Les interactions fortes avaient l'air très compliqué. 
Nous avons vu les efforts de les décrire en termes de propriétés de la matrice $S$ et nous avons mentionné le modèle de Veneziano. Il a suscité une grande activité théorique aussi bien pour comprendre ses propriétés mathématiques que pour développer des applications aux processus hadroniques. Dans ces efforts on trouve plusieurs contributions françaises\cite{Dual} qui ont continué durant la décennie suivante.  

Vers la fin des années soixante un nouvel accélérateur est entré en fonction à Stanford. C'était un accélérateur linéaire à électrons de plus de 3 km de longueur, atteignant une énergie de 50 GeV. Il a fourni une riche moisson de données concernant la diffusion d'électrons sur des nucléons -- protons ou neutrons -- à haute énergie de la forme $e^-+N\rightarrow e^-+X$, où $N$ désigne le nucléon de la cible et $X$ le système hadronique produit par la réaction.  La détection consistait à identifier l'électron final et à mesurer son impulsion. On se limitait aux événements à grand angle de diffusion pour lesquels l'électron transfère une grande quantité de mouvement au nucléon. On appelle cette région cinématique “profondément inélastique”.

L'intérêt de ce genre de réaction est facile à comprendre. L'électron n'a pas d'interaction forte, donc son interaction avec le nucléon est électromagnétique. C'est une interaction à laquelle nous pouvons appliquer la théorie des perturbations et la décrire en termes d'échange d'un photon virtuel qui va interagir avec les constituants chargés du nucléon. Le diamètre d'un nucléon est de l'ordre de 10$^{-13}$ cm. Un photon avec une énergie de 20 GeV a un pouvoir de résolution de l'ordre de 10$^{-15}$ cm, donc il peut “voir” des objets 100 fois plus petits et devient un excellent outil pour sonder l'intérieur du nucléon. L'image que nous avons obtenue a changé profondément notre perception des interactions fortes. 
Nous savons aujourd'hui que la complexité constatée auparavant ne montre qu'une partie superficielle de la réalité. Les interactions fortes apparaissent très compliquées parce que les objets qu'on utilise pour leur étude, à savoir les hadrons, sont eux-même compliqués. C'est comme si on essayait de découvrir les lois de l'électrodynamique en étudiant les interactions entre des macromolécules biologiques\footnote{Feynman donnait l'exemple suivant: imaginez que vous voulez étudier le mécanisme d'une montre suisse de grande précision et, pour le faire, vous en prenez deux et vous les cassez avec force l'une contre l'autre.}. 

Les résultats de Stanford étaient à la fois simples et surprenants. Le modèle de quarks était bien vérifié. Le nucléon apparaissait comme un “sac” contenant des quarks avec des charges électriques fractionnaires. Ce résultat était attendu. Celui qui ne l'était pas est que les quarks semblaient interagir avec le photon virtuel comme des particules libres, comme si ils n'étaient pas liés pour former le nucléon. Ceci était en contradiction avec le fait que la liaison était très forte parce qu'on n'arrivait pas à casser le nucléon et à libérer les quarks. Feynman avait proposé un modèle, dit “modèle des partons”, pour décrire ce comportement schizophrénique des quarks: l'intensité des forces qui lient les quarks à l'intérieur du nucléon dépendrait de la distance, très faible à courte distance, elle  deviendrait très forte à grande distance empêchant ainsi les quarks de se libérer. 

Ce modèle des partons était purement phénoménologique\footnote{On l'appelait souvent “modèle des partons na\"if”.} et l'effort pour lui donner un support  théorique solide s'est avéré passionnant et fructueux avec des contributions françaises\cite{Partons} intéressantes. L'aboutissement était une théorie quantique des champs pour les interactions fortes qui fait partie du Modèle Standard qu'on verra dans la section suivante.

\subsection{1970 - 1973 : LE MODELE STANDARD} 

\subsubsection{Les théories de Yang-Mills sont renormalisables.} En 1971 Gerard 't Hooft publia son premier article sur la renormalisation des théories de Yang Mills. Par la suite, dans une série d'articles entre 1971 et 1972, 't Hooft et Veltman ont démontré que les théories de Yang-Mills, aussi bien dans la phase symétrique avec les bosons de jauge de masse nulle, que dans celle de brisure spontanée avec des bosons massifs, sont renormalisables. Depuis cette date, les théories de jauge ont envahi la physique des particules élémentaires\footnote{S. Weinberg a décrit cette invasion avec “\dots and then, all hell broke loose”.}. 

La démonstration de 't Hooft et Veltman était basée sur l'analyse détaillée des diagrammes de Feynman avec des techniques inspirées de la théorie de la matrice $S$\footnote{En 1973 ils ont présenté un résumé de leur travail dans un rapport jaune du CERN sous le titre “Diagrammar”.}. B.W. Lee et J. Zinn-Justin ont publié une démonstration plus générale en utilisant les identités de Ward\cite{Lee-Zinn}. 

La formulation d'une théorie des champs invariante de jauge nécessite un choix de jauge qui brise l'invariance.  Nous obtenons ainsi une famille de théories, chacune ayant des propriétés très différentes et se pose la question de leur relation avec la théorie initiale.  Ce problème a été résolu par une équipe à Marseille sous la direction de Raymond Stora. Ils ont montré l'existence d'une nouvelle symétrie dont les transformations sont des fermions\cite{BRS}. Elle a énormément facilité la compréhension  des théories de jauge et elle est devenue un outil très puissant pour l'étude de tout autre système quantique avec contraintes. Des résultats similaires ont été obtenus par I. Tyutin en URSS et la symétrie est appelée “symétrie BRST” pour Becchi-Rouet-Stora-Tyutin. Elle a valu à ces chercheurs le prix Heineman de physique mathématique pour l'année 2009. 

\subsubsection{La théorie électrofaible.} Elle fut la première manifestation de la révolution suscitée par les théories de jauge. La propriété des théories de Yang-Mills spontanément brisées d'être renormalisables, donnait aux interactions faibles la puissance prédictive réservée jusqu'alors à l'électrodynamique. Le nombre de publications a explosé, en particulier sur les effets d'ordre supérieur.  En raison de l'absence de résultats expérimentaux, un grand nombre de modèles étaient aussi proposés, avec des propriétés diverses: avec ou sans courants neutres, avec brisure spontanée de la parité\cite{Fayet1} ou de $CP$ etc. Après la découverte des courants neutres par Gargamelle en 1973, ainsi que d'autres résultats expérimentaux qui suivirent et qui tendaient à confirmer le modèle de Weinberg-Salam, cette ligne de recherche s'est un peu estompée. 

Cependant, au début de 1972, un problème inattendu a surgi. La renormalisabilité des théories de Yang-Mills est basée sur les propriétés de symétrie qui nécessitent la conservation des courants. Or, comme on l'a déjà expliqué, les courants axiaux des interactions faibles contiennent des anomalies dans leurs équations de conservation. On s'est vite aperçu que ces anomalies rendaient le modèle initial de Weinberg-Salam non-renormalisable. Cette découverte\cite{BIM} a eu l'effet d'une douche froide. Heureusement, elle fut une courte alerte. La solution a été simple et physique. Le modèle de Weinberg-Salam est non-renormalisable parce qu'il contient uniquement les leptons. Si nous ajoutons les hadrons sous forme d'un modèle des quarks qui satisfait à la symétrie leptons-hadrons, comme dans le modèle du quark charmé, les anomalies entre leptons et hadrons s'annulent et le modèle complet est renormalisable\cite{BIM}.

La première confirmation éclatante de ce modèle vint en 1973 avec la découverte des courants neutres par Gargamelle. En 1974 Cl. et M.A. Bouchiat ont montré que les courants neutres des interactions faibles avaient des effets mesurables en physique atomique\cite{Bouchiat-2}. Ceci a ouvert un nouveau champ de recherche dans lequel l'équipe de M.A. Bouchiat à l'ENS a joué le rôle principal. 

\subsubsection{La liberté asymptotique et la chromodynamique quantique.} 
Les révolutions en Physique sont, le plus souvent, le résultat d'une découverte expérimentale inattendue.  Cela n'a pas été le cas avec la théorie standard électrofaible.  Aucune mesure n'était en contradiction avec la théorie de Fermi. La motivation était d'ordre esthétique.  Certains théoriciens cherchaient un schéma ayant une meilleure cohérence mathématique et une plus grande puissance prédictive. Pendant plusieurs années des prédictions de cette théorie -- les courants neutres, les particules charmées, les bosons vectoriels, le boson scalaire -- n'avaient aucune confirmation expérimentale. 

Par contre, la construction de la chromodynamique quantique, la théorie de jauge des interactions fortes, a été une réponse aux résultats de Stanford concernant la diffusion profondément inélastique d'électrons sur des nucléons. Il fallait donner un cadre théorique au modèle des partons que nous avons présenté à la section \ref{SLAC}. Nous ne présenterons pas cette histoire en détail parce qu'il n'y a pas eu de contribution importante de la part de chercheurs français. 

Le point de départ a été l'étude du groupe de renormalisation présenté à la section \ref{50-60-Int}. Il était connu depuis les années 50 mais son importance n'a été généralement comprise que vers la fin des années 60. Ces études ont montré que la force effective d'une interaction dépend de la distance et cette dépendance est décrite par les équations du groupe de renormalisation\footnote{Pour une explication simple voir les références \cite{LivreSFP} Chapitre 5, ainsi que \cite{Ilio1}.}.  

Pratiquement toutes les théories des champs renormalisables ont le comportement qu'on attend intuitivement: la force diminue avec la distance. Ce n'est pas celui que les expériences suggèrent pour la force entre les partons. En 1973 D.J. Gross et F. Wilczek à Princeton et indépendamment H.D. Politzer à Harvard, ont découvert que les théories de Yang-Mills dans la phase symétrique, c'est à dire avec les bosons de jauge de masse nulle, sont les seules à avoir un comportement contre-intuitif: la force augmente avec la distance. A très courte distance la force diminue et tend vers zéro. On appelle ce comportement “liberté asymptotique”. Il y a eu plusieurs tentatives d'expliquer ce phénomène de façon simple, mais aucune n'est vraiment satisfaisante. 

Gross, Wilczek et Politzer  ont postulé que les interactions fortes au niveau de quarks sont décrites par une théorie de jauge non brisée basée sur le groupe $SU(3)$ de la couleur\footnote{La propriété des théories de Yang-Mills d'être asymptotiquement libres avait été découverte en 1972 par G. 't Hooft, mais la connexion avec le modèle des partons n'avait pas été faite.}. C'est la “Chromodynamique Quantique.” Son accord avec les résultats expérimentaux, aussi bien des interactions lepton-nucléon que ceux des collisions hadroniques, est excellent. Il faut noter que la liberté asymptotique n'est pas valable si nous brisons la symétrie avec le mécanisme de Brout-Englert-Higgs. Par conséquent la chromodynamique quantique implique  des bosons de jauge de masse nulle.  $SU(3)$ en contient huit qui portent les charges de la couleur. On les appelle “gluons”. Afin d'expliquer le fait que nous n'avons jamais observé de gluons libres, nous supposons que, tout comme les quarks, ils sont “confinés” à l'intérieur des hadrons. Démontrer cette propriété de confinement pour la chromodynamique quantique reste un des grands problèmes non résolus de la théorie quantique des champs.

Avant de quitter cette section je voudrais mentionner une application du groupe de renormalisation aux phénomènes de transition de phase, initiée par K. Wilson. Dans ce cadre E. Brézin, J.C. Le Guillou et J. Zinn-Justin ont fait les calculs les plus précis des exposants critiques\cite{Gr-Ren}.

\subsection{\'Epilogue - Au delà du Modèle Standard}
Cet exposé se termine avec la construction du Modèle Standard comme la théorie de jauge du groupe $SU(3)\times SU(2) \times U(1)$ spontanément brisé en $SU(3)\times U(1)_{em}$. Bien sûr, l'histoire ne s'arrête pas en 1973. Il a fallu des décennies d'efforts, aussi bien de théoriciens que d'expérimentateurs, pour procéder à une comparaison détaillée de la théorie avec l'expérience. L'accord est spectaculaire. Il ne se limite pas aux grandes découvertes -- les courants neutres, les bosons vectoriels, les quarks $c, b$ et $t$, le boson de Brout-Englert-Higgs -- mais couvre aussi un vaste domaine de mesures  dont l'exposé  pourrait être le sujet d'une autre histoire passionnante. 

Feynman avait remarqué que {\it “\dots progress in physics is to prove yourself wrong as soon as possible.”} Notre problème est que, durant un demi siècle, le Modèle Standard a résisté à toutes les tentatives de le mettre en défaut. A l'inauguration de chaque nouvel accélérateur nous somme convaincus que {\it La Nouvelle Physique} sera enfin révélée, en vain jusqu'à présent. 

Nous avons plusieurs indications indirectes de l'existence d'une telle nouvelle physique, sans parler du fait que le Modèle Standard laisse des pans entiers de phénomènes (la matière noire, l'accélération de l'expansion de l'Univers, la gravitation etc) inexpliqués. Donc, depuis 50 ans, les théoriciens ont envisagé une pléthore de voies pour aller au delà du Modèle Standard. Le fil directeur était un préjugé théorique selon lequel la meilleure théorie est la plus symétrique. Par conséquent nous avons cherché à augmenter la symétrie du Modèle Standard. Les étapes successives faisaient appel à des théories mathématiques fascinantes sous des noms parfois exotiques --Théories Grandement Unifiées, Super-Symétrie, Super-Gravité -- mais aucune n'a reçu de confirmation expérimentale. 

 Un problème qui résiste à tous ces efforts est celui d'une théorie quantique de la gravitation. Même la théorie des champs avec la plus grande symétrie que nous pouvons imaginer, on l'appelle “super-gravité $N=8$”, ne semble pas capable de maîtriser toutes les divergences de la gravitation. Cet échec fut interprété comme un échec de la théorie quantique des champs. Le concept de champs quantiques locaux, qui traduit celui de particules ponctuelles, un concept hérité de la mécanique classique, fut abandonné. L'étape suivante fait appel à une théorie d'objets étendus. Le plus simple est uni-dimensionnel, une corde. La théorie  des cordes peut fournir une théorie quantique de la gravitation sans divergences, mais son contact avec l'expérience reste encore problématique. 
 
 Des physiciens français ont joué un rôle de premier plan dans la recherche d'une théorie au delà du Modèle Standard, la super-symétrie, la super-gravité ou la théories des cordes. Si ces efforts aboutissent, comme je l'espère, l'importance fondamentale de leurs travaux sera reconnue.

\section{Conclusions}

Raconter l'histoire du Modèle Standard serait une expérience exaltante. Les physiciens de ma générations ont eu la chance de vivre cette aventure extraordinaire, mais les acteurs  sont rarement des bons historiens. Des essais historiques existent déjà, voir par ex. \cite{Rise-SM}, mais une Histoire, au sens propre du terme, doit probablement attendre encore un peu. Dans cet exposé j'ai voulu profiter de l'anniversaire de la SFP pour parler des contributions françaises, même si je ne pouvais pas résister à la tentation de donner aussi un résumé de ma version de l'histoire. 

Quelles sont les conclusions:  un optimiste remarquera que, depuis la fin des années 50, l'importance des travaux français ne cesse d'augmenter, la dérivée est toujours positive; un pessimiste ne manquera pas de signaler la quasi-absence de la France durant plus de trente ans de l'histoire. Ce dernier point est troublant. L'histoire de la physique théorique des hautes énergies commence avec ce coup de génie de de Broglie et puis \dots rien. Je ne vois aucune autre grande nation scientifique qui présente un tel vide. 

Dans notre communauté il est courant de rendre de Broglie responsable de ce retard. Ce serait lui qui aurait empêché le développement et l'enseignement de la physique moderne en France. Ma compétence sur ce sujet est très limitée, néanmoins je trouve cette explication historiquement simpliste et, du point de vue pratique, contre-productive : trouver un bouc émissaire est un moyen commode pour ne pas chercher  des causes potentiellement dérangeantes. Je vais donner les raisons qui m'incitent à penser ainsi. 

D'abord, regardons la carrière de de Broglie\cite{Vila-Valls}. Il avait  hésité entre des études littéraire et scientifiques. Ayant opté pour les sciences, c'est seulement après la guerre qu'il a préparé une thèse sous la direction de Langevin. Il était personnellement fortuné et n'avait pas besoin d'un poste et d'un salaire. Il a travaillé comme bénévole dans le laboratoire de son frère Maurice de Broglie, laboratoire financé entièrement par  Maurice, jusqu'en 1928, année où il a obtenu un poste de Maître de Conférence à l'Université. C'était son premier poste; il avait 36 ans. L'année suivante, en 1929, il a reçu le prix Nobel de Physique {\it “for his discovery of the wave nature of electrons”}. Il avait prédit le phénomène de diffraction pour les électrons dans un de ses articles de 1923 et il semble qu'il avait essayé, sans succès,  de convaincre des expérimentateurs du laboratoire de son frère de faire l'expérience. Finalement cette prédiction fut confirmée en 1927 par diffusion d'un faisceau d'électrons à travers un cristal\footnote{Voir note en section \ref{MQ}. Pour cette observation G.P. Thomson et C.J. Davisson, les chefs de deux équipes, ont reçu le prix Nobel en 1937.}. Malgré le prix Nobel, de Broglie n'a été promu Professeur qu'en 1933, lorsque   la chaire de l'IHP est devenue vacante. Je ne pense pas qu'il existe un autre exemple de scientifique resté MdC pendant quatre ans avec un prix Nobel. 

L. de Broglie a occupé les postes scientifiques les plus prestigieux de France -- Secrétaire Perpétuel de l'Académie des Sciences, Membre de l'Académie Française -- mais il n'a jamais cherché à exercer un quelconque pouvoir “politique”, ou même un travail de “Direction”. Certes, il n'a pas utilisé son immense prestige pour défendre et diffuser la nouvelle physique, mais je n'ai pas l'impression qu'il ait empêché d'autres de le faire. Maurice Lévy, qui l'avait connu (de Broglie était membre de son jury de thèse), m'avait affirmé que de Broglie ne s'était jamais opposé à ses efforts pour reformer l'enseignement de la physique à l'Université. 
Dans un livre que de Broglie a publié en 1947\cite{deBroglie5} il dit avoir enseigné la mécanique quantique “officielle” pendant plusieurs années, mais je ne connais pas le contenu du cours ni le public à qui il était adressé. D'après Vila-Valls\cite{Vila-Valls}, ses cours étaient d'un niveau très élevé et destinés à des doctorants ou jeunes chercheurs. 
Si nous regardons sa liste de publications nous constatons qu'il est resté actif durant toute sa vie, même si le niveau de ses dernières publications est loin de celui de ses oeuvres de jeunesse. Mais reprocher à quelqu'un de n'avoir eu qu'une seule idée de génie, me paraît une critique de bien mauvaise foi. Depuis la fin des années 40 il s'est replié sur ses préoccupations pour une “nouvelle mécanique ondulatoire” et semble avoir perdu l'intérêt de ce qui se passait ailleurs. Ce comportement introverti n'est pas exceptionnel, on le trouve chez bien d'autres grands scientifiques vers la fin de leur carrière. 
En résumé, la vie scientifique de de Broglie, avec ses oscillations entre des périodes de gloire et de récession, ne me semble pas justifier l'image d'un dictateur qui bloque le développement de la science française. 

Alors, si ce n'est pas de Broglie, qui est le responsable? Je crains que la réponse ne soit “personne en particulier”; c'est l'échec de tout le système. Pour cette question les chapitres historiques du livre de la SFP\cite{LivreSFP} sont une mine précieuse d'informations. On y apprend que le problème est persistant et intemporel. Je cite quelques passages:

 Adolphe Wurtz, dans un rapport adressé au Ministère de l'Instruction publique,  note qu'en 1876 il y avait 293 étudiants inscrits dans l'ensemble des facultés des sciences françaises, nombre presque équivalent à celui des candidats au doctorat dans la Faculté des Sciences de la seule Université de Leipzig\footnote{Voir \cite{LivreSFP}, Chapitre 1.}. Dans un extrait d'un article de Pasteur de 1871, on lit : {\it “\dots la France s'est désintéressée, depuis un demi-siècle, des grands travaux de la pensée, particulièrement dans les sciences exactes.”}\footnote{Voir \cite{LivreSFP}, Chapitre 1.}.  J. Friedel, Président de la SFP en 1970, écrit pour la période entre les deux guerres: {\it “\dots deux normaliens par an seulement pouvaient commencer une recherche en physique, quand les promotions annuelles de jeunes docteurs au Cavendish\dots dépassaient la dizaine. En 1930, cinq cents étudiants suivaient les cours de mécaniques quantique à G\"ottingen\dots ”}\footnote{Voir \cite{LivreSFP}, Chapitre 3.}. Des chiffres qui font rêver. 

Pourquoi ce manque de soutien à la physique en général et à la physique théorique en particulier? Et pourquoi cette désaffection manifeste des étudiants Français pour les sciences? Plusieurs présidents de la SFP ont déploré {\it “\dots l'indifférence ou l'hostilité de la population\dots”}\footnote{H. Mathieu-Faraggi, janvier 1972, \cite{LivreSFP}, Chapitre 3.} vis-à-vis de la physique. Nous souhaitons attribuer cette attitude à un déficit de culture scientifique du public français, mais il me semble qu'on pourrait constater le même phénomène chez les scientifiques. Je me souviens d'un physicien qui se vantait en public de ne pas connaître la différence entre le Lagrangien et l'Hamiltonien. Une telle ignorance au pays de Lagrange \dots Mais il ne faisait que témoigner d'un fait avéré: il n'y avait pas que la mécanique quantique qui était mal enseignée, la mécanique classique l'était aussi. Lorsque j'enseignais à l'\'Ecole Polytechnique, les élèves entendaient le mot “Hamiltonien” uniquement au cours de mécanique quantique et je ne sais pas si ils se posaient la question de savoir pourquoi cet opérateur portait le nom d'un savant disparu bien des années avant la naissance de cette discipline. La seule et unique fois où le théorème de Noether fut enseigné à l'X de mon temps, était dans un cours de mathématiques de J.P. Bourguignon sous une forme abstraite, de sorte que nous, les physiciens, avions du mal à le reconnaître. 

Alors, une dernière question: pourquoi nous n'arrivons même pas aujourd'hui, à construire un programme d'enseignement de la physique cohérent et complet? Je vais risquer une réponse hérétique: parce que nous n'avons pas le temps. La France se vante d'être le pays de la Raison, mais, aux yeux d'un observateur extérieur, notre système d'enseignement de la physique semble défier toute analyse rationnelle. Je m'explique: dans tous les pays l'enseignement et en particulier celui de la physique, suit un programme standard. Les études secondaires sont suivies par celles à l'Université, en général entre 4 et 5 ans. C'est à ce stade que les étudiants apprennent la “physique moderne” par des enseignants-chercheurs. Ceux d'entre eux qui souhaitent compléter leur formation par la recherche intègrent une \'Ecole Doctorale et continuent souvent avec un stage post-doctoral. C'est le programme qui a fait ses preuves partout, sauf en France. 
Nous commençons par faire un tri à la fin du lycée: ceux qui ont des bonnes notes, surtout en mathématiques, sont choisis pour la filière d'élite destinée aux Grandes \'Ecoles, les autres peuvent s'inscrire à l'Université. Ces derniers auront la possibilité d'apprendre la physique par des chercheurs actifs, mais ils ne montrent pas toujours un tel intérêt. Les premiers, présumés plus doués si la sélection vaut quelque chose,  passent deux ans en classes préparatoires pendant lesquels ils n'ont aucun contact ni avec la science ni avec des scientifiques. Le programme de physique qu'ils suivent est très superficiel et s'arrête avant le milieu du XIXe siècle. Aucun sujet susceptible d'exciter la curiosité des jeunes n'est abordé\footnote{Je cite un collègue: “\dots concernant la physique, en secondaire et hypotaupe il m’avait semblé que le but était de décourager d’en faire\dots  C’est un peu comme le piano : on peut faire des gammes en continu, et quelques uns survivent. On peut aussi intéresser l’élève en exploitant très tôt la musicalité de pièces simples.”}. Les exigences du concours sont telles que les enseignants qui auraient envie de sortir un peu du programme imposé, et il y en a, sont activement découragés à le faire. Après le concours la grande majorité de ceux qui réussissent vont intégrer des écoles pour lesquelles la physique moderne est absente aussi bien de leur programme que de leur activité. Une petite minorité entre dans les écoles comme l'ENS ou l'X dans lesquelles les enseignants sont appelés à concocter un programme qui doit couvrir en deux ans ce qu'ailleurs on enseigne en quatre. Mission impossible, le temps manque. Pour résumer la logique: on applique un système de sélection probablement contestable et on offre un enseignement de physique correct uniquement à ceux que nous déclarons moins aptes à le suivre, logique qui laisse perplexes bien des observateurs extérieurs. 
Les physiciens français avec une brillante carrière n'ont pas réussi grâce au système d'enseignement mais malgré lui, avec l'aide d'initiatives comme l'\'Ecole des Houches. Ils offrent une illustration du dicton de E. Gibbon : {\it “The power of instruction is seldom of much efficacy, except in those happy dispositions where it is almost superfluous.”}


\begin{thebibliography}{99}
\bibitem{LivreSFP} “Les 150 ans de la Société Française de Physique” (2023), EDP Sciences SFP
\bibitem{Brillouin1}  Marcel Brillouin (1919), CRAS 168, p. 1318
\bibitem{Pestre} D. Pestre (1984), “Physique et physiciens en France 1918-1940”, deuxième édition, 1992, Paris, éditions des
archives contemporaines. {\it Voir aussi:}
(1985),  “Y a-t-il eu une physique à la française entre les deux guerres ?”, La recherche, 169.
septembre 1985, pp. 999-1005. 
\bibitem{Vila-Valls} A. Vila-Valls (2012), “Louis de Broglie et la diffusion de la mécanique
quantique en France (1925-1960)”, Thèse, Université Claude Bernard Lyon 1, https://tel.archives-ouvertes.fr/tel-00993036/document.
\bibitem{Pais} A. Pais (1988), “Inward Bound”, Oxford University Press.
\bibitem{deBroglie0} L. de Broglie (1972), Allocution prononcée à l'occasion de son 80ème anniversaire, Archives de l’Académie des sciences.
\bibitem{deBroglie1} L. de Broglie (1922), Journal de Physique, 6e série, t. III, p. 422.
\bibitem{deBroglie2} L. de Broglie (1923), CRAS 177, p. 507.
\bibitem{deBroglie3} L. de Broglie (1923), CRAS 177, p. 548 et p. 630. Voir aussi L. Brillouin, (1924), CRAS 178, p. 1696.
\bibitem{deBroglie31} L. de Broglie (1924), CRAS 179, p. 39 ;  Phil. Mag. 47, p. 446.
\bibitem{Bril1} L. Brillouin (1926), CRAS 183, p.24.
\bibitem{deBr32} L. de Broglie (1927), J. de Phys. et Le Rad. 8, p. 225.
\bibitem{Solvay27} CR Congrès Solvay 1927, {\it “Electrons et photons, Rapports et discussions du Conseil de physique tenu à Bruxelles du 24 au 29 Octobre 1927”}, Paris (1928), Gauthier-Villars.
\bibitem{Lev-Lebl} J.M. Lévy-Leblond (1967), Comm. Math. Phys. 6, p. 286.
\bibitem{Brill} L. Brillouin (1930), CRAS 191, p. 292.
\bibitem{Perrin1} F. Perrin (1933), CRAS 197, p. 1625.
\bibitem{deBroglie4} L. de Broglie (1934), CRAS 198, p. 135.
\bibitem{Proca1} A. Proca (1936), J. de Phys. et Le Rad. 7, p. 347.
\bibitem{LePrince} L. Leprince-Ringuet, M. Lhéritier (1944), CRAS 219, p. 618.
\bibitem{Froissart} M. Froissart (1961), Phys. Rev. 123, p. 1053.
\bibitem{Jacob-W} M. Jacob, G. C. Wick (1959), Annals Phys. 7, p. 404.
\bibitem{Levy1} M. Lévy (1952), Phys. Rev. 88, p. 725.
\bibitem{Levy2} M. Gell-Mann, M. Lévy (1960), Nuov. Cim. 16, p. 605
\bibitem{Levy3} Pour l'oeuvre de Lévy voir les témoignages recueillis à l'occasion de son 90e anniversaire : https://maurice-levy-physicien.fr
\bibitem{Michel1} L. Michel (1949), Nature 163, p. 959 ; L. Michel (1950), Proc. Phys. Soc. London A63, p. 514 et p. 1371. 
\bibitem{Michel2} Cl. Bouchiat, L. Michel (1957), Phys. Rev. 106, p. 170.
\bibitem{Michel21} Cl. Bouchiat, L. Michel (1961), J. de Phys. et Le Rad. 22, p. 121.
\bibitem{Michel3} L. Michel (1953), Nuov. Cim. 10, p. 319.
\bibitem{Michel4} V. Bargmann, L. Michel, V.L. Telegdi (1959), Phys. Rev. Lett. 2, p. 435.
\bibitem{Bouchiat1} Cl. Bouchiat (1958), Phys. Rev. Lett. 1, p. 351 ; Phys. Rev. 112, p. 877.
\bibitem{CERN1} L'histoire de la création du CERN, ainsi que ses premières années de fonctionnement, sont présentées dans “Studies in CERN history” (1993), CERN. Un Chapitre spécial raconte l'histoire de la physique théorique.
\bibitem{Esp-Pr1} B. d'Espagnat, J. Prentki (1955), Phys. Rev. 99, p. 328 ; B. d'Espagnat, J. Prentki, A. Salam (1957), Nucl. Phys. 3, p. 446 ; (1958), Nucl. Phys. 5, p. 447 ; B. d'Espagnat, J. Prentki (1958), Nucl. Phys. 6, p. 596.
\bibitem{Esp-Pr2} B. d'Espagnat, J. Prentki (1962), Nuov. Cim. 24, p. 497. 
\bibitem{Omnes} R. Omnes (1958), Nuov. Cim. 8, p. 316.
\bibitem{Br-Ep-Gl} J. Bros, H. Epstein, V. Glaser (1964), Nuov. Cim. 31, p. 1265 ;  (1965), Comm. Math. Phys. 1, p. 240 ;  (1972), Helv. Phys. Acta 45, p. 149 ; H. Epstein, V. Glaser, A. Jaffe (1965), Nuov. Cim. 36, p. 1016 ; J. Bros, H. Epstein, V. Glaser, M. Danowski (1967), Comm. Math. Phys. 6, p. 77.
\bibitem{Martin} A. Martin (1969), {\it “Scattering Theory. Unitarity, Analyticity and Crossing”} Lecture notes in Physics, Vol. 3, Springer-Verlag ; H. Epstein, V. Glaser, A. Martin (1969), Comm. Math. Phys. 13, p. 257.
\bibitem{Diu-Rub} B. Diu, J.L. Gervais, H. Rubinstein (1963 ), Phys. Lett. 4, p. 280 ;  (1964), Nuov. Cim. 31, p. 27  et p. 341.
\bibitem{COMONAV} G. Cohen-Tannoudji, A. Morel, H. Navelet (1968), Annals Phys. 46, p. 239.
\bibitem{Bacry-N-VH} H. Bacry, J. Nuyts, L. Van Hove (1964), Phys. Lett. 9, p. 279.
\bibitem{Orsay-quarks} A. le Yaouanc, L. Oliver, O. Pene, J.C. Raynal (1973), Phys. Rev. D8, p. 2223.
\bibitem{Itz-etal} B. Diu (1963), Nuov. Cim. 28, p. 466 ; Cl. Itzykson, M. Nauenberg (1966), Rev. Mod. Phys. 38, p. 95 ; M. Bander, Cl. Itzykson (1966), Rev. Mod. Phys. 38, p. 330 et  p. 346 ; H. Bacry, J.M. Levy-Leblond (1968), J. Math. Phys. 9, p. 1605. 
\bibitem{Bouch-Mey1} Cl. Bouchiat, Ph. Meyer (1964), Nuov. Cim. 34, p. 1122.
\bibitem{Gell-Mich} J. Bernstein, M. Gell-Mann, L. Michel (1960), Nuov. Cim. 16,  p. 560.
\bibitem{Fr-St} M. Froissart, R. Stora (1960), Nucl. Instr. Meth. 7, p. 297.
\bibitem{Stora1} C. De Callan, R. Stora, W. Zimmermann (1969), Nuov. Cim. Lett. 1, p. 877.
\bibitem{Itzyk2} E. Brézin, Cl. Itzykson, J. Zinn-Justin (1970), Phys. Rev. D1, p. 2349 ; E. Brézin, Cl. Itzykson (1970), Phys. Rev. D2, p. 1191 ;  (1971), Phys. Rev. D3, p. 618.
\bibitem{deRaf1} B.E. Lautrup, A. Peterman, E. De Rafael (1970), Phys. Lett. B31, p. 577.
\bibitem{deRaf} B.E. Lautrup, E. De Rafael (1968), Phys.Rev. 174, p. 1835 ; M. Gourdin, E. De Rafael (1969), Nucl.Phys.B 10, p. 667 ; J.S. Bell, E. de Rafael (1969), Nucl.Phys.B 11, p. 611 ; B.E. Lautrup, E. de Rafael (1969), Nuovo Cim. 64, p. 322 ;  B.E. Lautrup, A. Peterman, E. De Rafael (1971) Nuovo Cim. A1 p. 238. 
\bibitem{Kastler1} R. Haag, D. Kastler (1964), Journ. Math. Phys. 5, p. 848.
\bibitem{Itzyk1} H. D. I. Abarbanel, Cl. Itzykson (1969), Phys. Rev. Lett. 23, p. 53.
\bibitem{Epst-Gl} H. Epstein, V. Glaser (1973), Ann. Inst. H. Poincaré A19, p. 211.
\bibitem{Stora2} H. Epstein, V. Glaser, R. Stora (1973), Lect. Notes Math. 449, p. 143 ; R. Stora (1973), Lect. Notes Math. 449, p. 163 ; J. Bros, H. Epstein, V. Glaser, R. Stora (1973), Lect. Notes Math. 449, p. 185.
\bibitem{Amati1} D. Amati, Cl. Bouchiat, J.L. Gervais (1970), Nuov. Cim. A65, p. 55.
\bibitem{SSB1} P. Curie (1894), Journ. de Phys. théor. et appl. 3, p. 393 ; P.G. de Gennes (1981) IDSET-Paris p. 1.   
\bibitem{Kastler2} D. Kastler, D.W. Robinson, A. Swieca (1966), Comm. Math. Phys. 2, p. 108.
\bibitem{Aspect} A. Aspect, P. Grangier, G. Roger (1982), Phys. Rev. Lett. 49, p. 91 ; voir aussi A. Aspect (1991), Europhys. News 22, p. 73.
\bibitem{Rise-SM} L. Brown {\it et al} (1997), {\it “The Rise of the Standard Model”}, Cambridge University Press ; L. Maiani, L. Rolandi eds. (2016), {\it “The Standard Theory of Particle Physics”}, World Scientific.
\bibitem{Ilio1} J. Iliopoulos (2015), {\it “Aux origines de la masse. Particules élémentaires et symétries fondamentales.”}, Editions de Physique.
\bibitem{BIP} Cl. Bouchiat, J. Iliopoulos, J. Prentki (1968),  Nuov. Cim. 56A, p. 1150 ; J. Iliopoulos (1969), Nuov. Cim. 62A, p. 209.
\bibitem{GIM} S.L. Glashow, J. Iliopoulos, L. Maiani (1970), Phys. Rev. D2, p. 1285.
\bibitem{Amati2} D. Amati, H. Bacry, J. Nuyts, J. Prentki (1964), Phys. Lett. 11, p. 190 ; Nuov. Cim. 34, p. 1732.
\bibitem{Dual} D. Amati, Cl. Bouchiat, J.L. Gervais (1969), Nuov. Cim. Lett. 2, p. 399 ; Cl. Bouchiat, J.L. Gervais, N. Sourlas (1970), Nuov. Cim. Lett. 3, p. 767.
\bibitem{Partons} H. Leutwyler, J. Stern (1970), Phys. Lett. B31, p. 458 ;  Nucl. Phys. B20, p. 77 ; Cl. Bouchiat, P. Fayet, Ph. Meyer (1971), Nucl. Phys. B34, p. 157 ;  Cl. Bouchiat, P. Fayet, N. Sourlas (1972), Nuov. Cim. Lett. 4, p. 9.
\bibitem{Lee-Zinn} B.W. Lee, J. Zinn-Justin (1972), Phys. Rev. D5, p. 3121 ; p. 3137 ; p. 3155 ;  (1973) Phys. Rev. D7, p. 1049.
\bibitem{BRS} A. Rouet, R. Stora (1972), Nuov. Cim. Lett. 4, 136 ; L. Quaranta, A. Rouet, R. Stora, E. Tirapegui (1972), Nuov. Cim. Lett. 4, P. 924 ; C. Becchi, A. Rouet, R. Stora (1974), Phys. Lett. B52, p. 344 ;  (1975), Comm. Math. Phys. 42 p. 127 ;  (1975), Annals Phys. 98, p. 287.
\bibitem{Fayet1} P. Fayet (1974), Nucl. Phys. B78, p. 14.
\bibitem{BIM} Cl. Bouchiat, J. Iliopoulos, Ph. Meyer (1972), Phys. Lett. B38, p. 519.
\bibitem{Bouchiat-2} M.A Bouchiat, Cl. Bouchiat (1974), Phys. Lett. B48, p. 111 ;  Journal de Phys. 35, p. 899.
\bibitem{Gr-Ren} E. Brézin, J.C. Le Guillou, J. Zinn-Justin (1973), Phys. Rev. D8, p. 434 ; p. 2418 ; (1974), Phys. Rev. D9, p. 1121 : D10, p. 2046.
\bibitem{deBroglie5} L. de Broglie (1947), {\it “Physique et microphysique”}, Paris, Albin Michel 
\end{thebibliography}
\end{document}